\documentclass[a4paper,12pt]{article}

\usepackage{booktabs,dcolumn}
\usepackage{psfrag,graphicx}
\usepackage{eso-pic,graphicx}
\usepackage{amsmath}
\usepackage{graphicx}
\usepackage{pdflscape}
\usepackage{longtable}
\usepackage{epstopdf}
\usepackage{appendix}
\usepackage{dashbox}
\usepackage{hyperref}
\usepackage{natbib}

\usepackage[utf8]{inputenc}
\usepackage{natbib}
\usepackage{amsmath}
\usepackage{dashbox}
\usepackage{appendix}
\usepackage{amsmath}
\usepackage{amsfonts}
\usepackage{amssymb}
\usepackage{subfig}
\usepackage{psfrag}
\usepackage{graphicx}
\usepackage{epsfig}
\usepackage{float}
\usepackage{bm}
\usepackage{booktabs}
\usepackage{multirow}
\usepackage{subfiles}
\usepackage{amsbsy,enumerate}
\usepackage{natbib}
\usepackage{pgffor}
\usepackage{tabularx,colortbl}
\usepackage{tikz}
\usepackage{cancel}

\usepackage [ normalem ] { ulem }

\newcolumntype{z}[1]{D{.}{.}{#1}}

\newcommand{\cred}{\textcolor{red}}

\begin{document}

\title{

\begin{center}
{\Large \bf A semi-parametric dynamic conditional correlation framework for risk forecasting} \end{center}
}


\author{Giuseppe Storti$^{1}$\footnote{Email: storti@unisa.it.}, Chao Wang$^{2}$\footnote{Corresponding author. Email: chao.wang@sydney.edu.au.}
\\
$^{1}$ Department of Economics and Statistics, University of Salerno\\
$^{2}$ Discipline of Business Analytics, The University of Sydney}
\date{} \maketitle

\begin{abstract}
\noindent

We develop a novel multivariate semi-parametric framework for joint portfolio Value-at-Risk (VaR) and Expected Shortfall (ES) forecasting. Unlike existing univariate semi-parametric approaches, the proposed framework explicitly models the dependence structure among portfolio asset returns through a dynamic conditional correlation (DCC) parameterization. To estimate the model, a two-step procedure based on the minimization of a strictly consistent VaR and ES joint loss function is employed. This procedure allows to simultaneously estimate the DCC parameters and the portfolio risk factors. The performance of the proposed model in risk forecasting on various probability levels is evaluated by means of a forecasting study on the components of the Dow Jones index for an out-of-sample period from December 2016 to September 2021. The empirical results support effectiveness of the proposed framework compared to a variety of existing approaches.

\end{abstract}

\newpage
\pagenumbering{arabic}

{\centering
\section{ Introduction}\label{introduction_section}
\par
}
\noindent
Value-at-Risk (VaR) and Expected Shortfall (ES) play a central role in the risk management systems of banks and other financial institutions. For more than two decades, VaR has been the official risk measure adopted worldwide by financial intermediaries operating in the global financial system. Nevertheless, VaR has some important theoretical limits. First, VaR cannot measure the expected loss for extreme (violating) returns. In addition, it can be shown that VaR is not always a \emph{coherent} risk measure, due to failure to match the \emph{subadditivity} property.  For these reasons, the Basel Committee on Banking Supervision proposed in May 2012 to replace VaR with the ES (\citealt{artzner1997,artzner1999}). The prominent role of ES in the new international risk measurement practice is clearly illustrated by the official standards for determining minimum capital requirements for market risk published on the BIS institutional website \citep[see e.g. section MAR33]{BIS2019std} and effective from January 2022. Thus, in recent years ES has been increasingly employed for tail risk measurement. Analytically, ES is defined as the expectation of the return conditional on the VaR being exceeded and, unlike VaR, it is a  coherent measure and ``measures the riskiness of a position by considering both the size and the likelihood of losses above a certain confidence level'' \citep{BIS2013}.

Our paper focuses on the daily forecasting of VaR and ES on the lower/left tail. In line with the Basel III Accord and studies in the literature, our paper analyses the commonly used $\alpha=1\%$, 2.5\% and 5\% probability levels.

Compared to VaR, there is much less existing work on modeling ES, which is partly due to the non-elicitability of ES alone. However, the recent work in \cite{Fissler2016} has shown that the pair (VaR,ES) is jointly elicitable. They develop a family of joint loss, or \emph{scoring}, functions that are strictly consistent for the true VaR and ES, that is, they are uniquely minimized by the true VaR and ES series. This result has important implications for the estimation of conditional VaR and ES as well as for ranking risk forecasts from alternative competing models.

\cite{tayl2019} proposes a joint VaR and ES modelling approach (named as ES-CAViaR in this paper) based on the minimization of the negative of an Asymmetric Laplace (AL) log-likelihood function that can be derived as a special case of the \cite{Fissler2016} class of loss functions, under specific choices of the functions involved. Furthermore, \cite{pattonetal2019} propose new
dynamic models for VaR and ES, through adapting the generalized autoregressive score (GAS) framework \citep{crealetal2013} and using a 0-degree homogeneous loss function that falls in the \cite{Fissler2016} class (FZ0).

The works mentioned above focus on univariate time series and do not take into account the correlation among assets in financial markets. Several quantile-based methods, see for example the works in \cite{baur2013structure}, \cite{bernardi2015bayesian}, \cite{white2015var}, have been developed to estimate VaR in a multivariate setting and model the tail interdependence between assets, while the ES component is not specified in these frameworks.

 We contribute to the literature on portfolio risk forecasting by proposing a novel class of semi-parametric multivariate GARCH (MGARCH) models that jointly generate portfolio VaR and ES forecasts, using multivariate information on the returns of portfolio constituents as input. Our main reference model is given by a semi-parametric version of the dynamic conditional correlations (DCC) model \citep{engle2002dynamic} via proposing a Semi-DCC framework. Its parameters are estimated by minimizing a strictly consistent joint VaR and ES loss function that fits into the FZ class of \cite{Fissler2016}, without requiring the formulation of restrictive assumptions on the conditional distribution of portfolio returns.

The proposed framework can be easily extended to other parameterizations from the multivariate GARCH literature such as the corrected DCC (cDCC) of \cite{aielli2013dynamic} or dynamic covariance models such as the BEKK model of \cite{engle1995multivariate}. Moreover, the Semi-DCC framework could be conveniently extended to more complex DCC parameterizations that allow for asymmetries and the inclusion of exogenous variables, such as the Asymmetric Generalized DCC (AG-DCC) family of models proposed by \cite{cappiello2006asymmetric}.

There are two different ways to look at our proposed framework. Namely, the Semi-DCC can be viewed either as a semi-parametric DCC model or as an ES-CAViaR for portfolio risk forecasting. It should be noted, however, that compared to each of these approaches, the Semi-DCC is characterized by some distinctive and innovative features.

In fact, compared to the standard DCC, the structure of the Semi-DCC is made up of three components, of which the first two, relating to the specification of volatility and correlation dynamics, are common to the standard DCC model, while the third, relating to portfolio risk estimation, is specific to our model. We refer to our proposed model as a semi-parametric DCC model, since the risk model, while still using multivariate information on asset correlations, is fitted to the time series of univariate portfolio returns $r_t(\mathbf{w})$, for a given choice of portfolio weights $\mathbf{w}$. Furthermore, in the DCC model the dynamic conditional correlation parameters are usually estimated by Gaussian Quasi Maximum Likelihood (DCC-QML) through the maximization of a multivariate normal quasi-likelihood function. Then the portfolio VaR and ES could be ex-post obtained through the filtered historical simulation (HS) approach by calculating the sample quantiles and tail averages of the standardized returns \citep{francq2015risk}. In contrast, the Semi-DCC takes a more natural approach to VaR and ES estimation by fitting the risk model directly to the time series of portfolio returns rather than to the standardized returns. Specifically, the Semi-DCC involves the optimization of a univariate risk-targeted strictly consistent loss function for portfolio VaR and ES, e.g. the AL or FZ0 joint loss function, allowing portfolio VaR and ES to be estimated directly. An additional strength of this approach is that it also allows for the simultaneous estimation of dynamic correlation parameters and portfolio risk factors, whereas in standard DCC models these are estimated in two subsequent steps. The semi-nonparametric (SNP) DCC model proposed by \cite{delbrio2011} and \cite{niguez2016} also replace the second-stage Gaussian QML with more flexible alternatives based on series expansions. Although this class of models is similar in spirit to ours in attempting to provide a more flexible alternative to the standard DCC, the Semi-DCC model differs from the SNP-DCC in two main ways. First, the Semi-DCC model is fitted to the series of univariate portfolio returns. Second, in the SNP-DCC the second stage estimation is based on an approximate likelihood that is not ``risk-targeted'' in the sense of being strictly consistent for the pair (VaR, ES).

Compared to the standard ES-CAViaR framework, the Semi-DCC model extends it to a multivariate setting, where it allows for joint estimates of portfolio VaR and ES that explicitly take into account the cross-sectional correlation structure of portfolio returns. From a risk forecasting perspective, the Semi-DCC model can then be viewed as an ES-CAViaR that takes multivariate input information on the asset variance and covariance matrix, which is then used to obtain univariate VaR and ES measures for portfolio returns. However, unlike univariate semi-parametric approaches to risk forecasting, the underlying multivariate GARCH structure allows the Semi-DCC model to be used for a wider range of applications, including portfolio allocation and hedging.

\cite{merlo2021} have recently proposed a multivariate model for generating joint forecasts of the pair (VaR, ES). Their framework is based on a multivariate extension of the AL distribution and allows for portfolio risk forecasting and optimization. However, their approach is limited to low portfolio dimensions (three market indices are studied in their paper) and is not semi-parametric, since inference is based on
maximum likelihood estimation and not on the minimization of a
strictly consistent scoring function.

The empirical performance of the Semi-DCC using AL and FZ0 joint losses model is assessed by means of an application to a panel of 28 assets included in the Dow Jones index. Our findings can be summarized as follows. When forecasting risk for an equally weighted portfolio, Semi-DCC models are competitive with state-of-the-art univariate semi-parametric approaches and perform better than conventional DCC models. Further, we assess the performance of the proposed model in constructing global minimum ES portfolios. Under this respect, our results show that the Semi-DCC model clearly outperforms a benchmark equally weighted allocation strategy. We also find evidence that in a turbulent period centred on the outbreak of the COVID-19 crisis in 2020, the Semi-DCC model leads to less risky portfolios compared to existing DCC specifications.

The paper is structured as follows. Section \ref{methodology_section}, after providing a brief  review of the existing semi-parametric univariate risk forecasting approaches, presents the proposed class of semi-parametric DCC models, while their estimation procedure is illustrated in Section \ref{estimation_section}. The finite sample properties of the estimators and the risk forecasting performance of Semi-DCC are investigated in Section \ref{simulation_section} via an extensive Monte Carlo simulation study. Section \ref{empirical_section} discusses the results of an empirical application to risk forecasting for the constituents of the Dow Jones index, where the portfolio optimization results are included in Appendix \ref{po_empirical} to limit the focus of the empirical study. Section \ref{conclusion_section} concludes the paper.


\section{Statistical framework}\label{methodology_section}

\subsection{Description of the environment}
\label{model_proposed_section}
 First, let $\mathbf{r}_t=(r_{t,1},\ldots,r_{t,n})'$ be the vector of log-returns on $n$ portfolio assets at time $t$. Further assume that $\mathbf{r}_t$ is generated by the process:
\begin{equation}
 \mathbf{r}_t=\boldsymbol{\mu}_t+H^{1/2}_t \mathbf{z}_t,
 \label{e:obs}
\end{equation}
where $\mathbf{z}_t\overset{iid}{\sim} D_1(\mathbf{0},I_n)$, $D_1$ is a multivariate distribution with zero mean and identity covariance matrix, $\boldsymbol{\mu}_t=E(\mathbf{r}_t|\mathcal{I}_{t-1})$ is the conditional mean vector of returns with $\mathcal{I}_{t-1}$ as the information available at time $t-1$, $H^{1/2}_t$ is a positive definite matrix such that $H^{1/2}_t (H^{1/2}_t )'=H_t$ and $H_t=\mathbb{V}(\mathbf{r}_t|\mathcal{I}_{t-1})$ is the conditional variance and covariance matrix of returns, with  $\mathbb{V}$ being the variance operator.

Pre-multiplying both members of Equation (\ref{e:obs}) by the (transposed) vector of portfolio weights $\mathbf{w}=(w_1,\ldots,w_n)'$, we obtain the (univariate) portfolio returns:
\begin{equation}
    r_t(\mathbf{w})=\mathbf{w}^{'}\mathbf{r}_t=\mathbf{w}^{'}\boldsymbol{\mu}_t+\mathbf{w}^{'}{H}^{1/2}_t\mathbf{z}_t. \nonumber
    \label{e:locscale0}
\end{equation}
It is easy to infer that these can be equivalently represented as:
\begin{equation}
    r_t(\mathbf{w})=\mathbf{w}^{'}\boldsymbol{\mu}_t+{e}_t(\mathbf{w})\sqrt{\mathbf{w}^{'}{H}_t\mathbf{w}},
    \label{e:locscale1}
\end{equation}
where ${e}_t(\mathbf{w})$ is a scalar continuous error term such that ${e}_t(\mathbf{w})\overset{iid}{\sim}(0,1)$. Assuming that portfolio returns follow the process in  (\ref{e:locscale1}) and given the set of weights $\mathbf{w}$, the $\alpha$-level portfolio conditional VaR and ES are given by:
\begin{eqnarray} \label{e:modvar_es_1}
Q_{t,\alpha}(\mathbf{w})=\mathbf{w}^{'}\boldsymbol{\mu}_t + q_{\alpha}(\mathbf{w}) \sqrt{\mathbf{w}^{'}{H}_t\mathbf{w}},  \qquad
\text{ES}_{t,\alpha}(\mathbf{w})=\mathbf{w}^{'}\boldsymbol{\mu}_t +  c_{\alpha}(\mathbf{w}) \sqrt{ \mathbf{w}^{'}{H}_t\mathbf{w}},
\end{eqnarray}
where $\mathbf{w}^{'}\boldsymbol{\mu}_t$ and $\mathbf{w}^{'}{H}_t\mathbf{w}$ are the conditional mean and variance of portfolio returns $r_t(\mathbf{w})$, respectively, and $\alpha$, with $0<\alpha<<1$, denotes the target level for the estimation of VaR and ES. Furthermore, letting $F_{e_t(\mathbf{w}),t-1}(.)$ represent the Cumulative Distribution Function (CDF)  of $e_t(\mathbf{w})$ conditional on $\mathcal{I}_{t-1}$ and assuming that this is continuous on the real line, we define $q_{\alpha}(\mathbf{w})=F^{-1}_{e_t(\mathbf{w}),t-1}(\alpha)$ and $ c_{\alpha}(\mathbf{w})=E(e_t(\mathbf{w})|\mathcal{I}_{t-1},e_t (\mathbf{w}) \le q_{\alpha}(\mathbf{w}) )$. It it is worth noting that, when the innovations $\mathbf{z}_t$ are assumed to have a spherical distribution (implying that the distribution of $e_t(\mathbf{w})$ will be the same as the marginal distribution of the components of $\mathbf{z}_t$), the values of $q_{\alpha}(\mathbf{w}) $ and $c_{\alpha}(\mathbf{w})$ will not be dependent on the portfolio weights $\mathbf{w}$. Also,  assuming that the innovations $\mathbf{z}_t$ in
(\ref{e:obs}) have a spherical distribution, implies that the conditional distribution of returns $\mathbf{r}_t$ belongs to the elliptical family  \citep[see e.g.,][]{frazak2020}.

In the remainder of Sections \ref{methodology_section} and \ref{estimation_section}, considering that we are studying risk forecasting at a given probability level (1\%, 2.5\% or 5\%) for a pre-defined portfolio composition $(\mathbf{w})$, the following notational conventions will be adopted to simplify the presentation:  $Q_{t,\alpha}(\mathbf{w}) \equiv Q_t$, $\text{ES}_{t,\alpha}(\mathbf{w})\equiv \text{ES}_t$, $q_{\alpha}(\mathbf{w})\equiv q$, $c_{\alpha}(\mathbf{w})\equiv c$ and $r_t(\mathbf{w}) \equiv r_t$.


\subsection{Univariate semi-parametric approaches to portfolio risk forecasting} \label{background_section}

The literature on semi-parametric forecasting of portfolio risk, VaR and ES, has so far mostly been limited to univariate approaches. In this section, we present a selective review of the most relevant work on this topic.

Focusing on VaR forecasting, \cite{koenkermachado1999} note that the usual quantile regression estimator is equivalent to a maximum likelihood estimator based on the AL density with a mode at the quantile. The parameters in the model for $Q_t$ can then be estimated maximizing a quasi-likelihood based on:
$$ p(r_t| \mathcal{I}_{t-1}) = \frac{\alpha (1-\alpha)}{\sigma} \exp \left( -\frac{(r_t-Q_t)(\alpha - I(r_t \le Q_t) )} {\sigma}  \right), $$
for $t=1,\ldots,N$ and where $\sigma$ is a scale parameter.

\cite{tayl2019}, noting a link between $\text{ES}_t$ and a dynamic $\sigma_t$, extends this result to incorporate the associated ES quantity into the likelihood expression, resulting in the conditional density function:
\begin{eqnarray} \label{e:es_var_likelihood}
p(r_t| \mathcal{I}_{t-1}) = \frac{(\alpha - 1)}{\text{ES}_t} \exp \left( \frac{(r_t-Q_t)(\alpha - I(r_t \le  Q_t)) }{\alpha \text{ES}_t}  \right).
\end{eqnarray}

This allows a quasi-likelihood function to be built and maximised, given model expressions for $(Q_t, \, \text{ES}_t)$. \cite{tayl2019} supports the validity of this estimation procedure noting that the negative logarithm of the resulting likelihood function is strictly consistent for $(Q_t,\, \text{ES}_t)$ considered jointly, that is it fits into the FZ class of jointly consistent scoring functions for VaR and ES developed by \cite{Fissler2016}.

Along the same lines, \cite{pattonetal2019} investigate a class of semi-parametric models, including some observation driven models, whose parameters can be estimated minimizing a 0-degree homogeneous loss function (FZ0) still included in the same FZ class, assuming that both VaR and ES are both strictly negative. The FZ0 loss used in \cite{pattonetal2019} can be shown as:
\begin{equation}
FZ0=-\frac{1}{\alpha \text{ES}_t}I(r_t \le  Q_t) \,(Q_t-r_t)+\frac{Q_t}{\text{ES}_t}+\log(-\text{ES}_t)-1
\label{e:fz0_loss_original}
\end{equation}

As mentioned above, all these papers focus on univariate semi-parametric modeling approaches that, when applied to portfolio returns, do not explicitly assess the impact of cross-sectional correlations among assets. Although this issue has been extensively analyzed in the literature on parametric MGARCH models \citep{bauwensetal2006}, there is much less existing work in addressing the issue in a multivariate semi-parametric risk-targeted framework. This motivates the proposal of a novel semi-parametric DCC modeling strategy.

\subsection{A semi-parametric DCC framework}
\label{ss:spmgarch}

The analytical expressions for portfolio VaR and ES provided in Equation (\ref{e:modvar_es_1}) make evident that the VaR and ES dynamics are driven by those of the conditional variance and covariance matrix $H_t$. In this section we describe a semi-parametric framework for assessing the impact that conditional correlations among portfolio components and their individual volatilities have on portfolio risk forecasts.

The dynamics of $H_t$ can be modelled using a wide range of specifications from the MGARCH literature. Without any loss of generality, due to its flexibility and widespread diffusion among practitioners, our proposed modeling approach builds on the DCC model of \cite{engle2002dynamic} and shares the same volatility and correlation dynamics. However, compared to the standard DCC, our framework incorporates an additional step related to the specification of portfolio VaR and ES.

In particular, the risk model is fitted to the time series of univariate portfolio returns $r_t(\mathbf{w})$, for a given set of portfolio weights $\mathbf{w}$. The \emph{semi-parametric} nature of the model derives from the fact that, as it will be later discussed in Section \ref{estimation_section}, its coefficients are fitted minimizing a jointly consistent loss for $Q_t$ and $\text{ES}_t$, e.g., AL (Equation (\ref{e:es_var_likelihood})) or FZ0 (Equation (\ref{e:fz0_loss_original})), without assuming the parametric distribution of returns.

Under the DCC specification, the conditional variance and covariance matrix $H_t$ is decomposed as:
\begin{equation}
H_t= D_t P_t D_t,
\label{e:dcc1}
\end{equation}
where $D_t$ is a $(n \times n)$ diagonal matrix such that its i-th diagonal element is $D_{t,ii}=h_{t,i}$, with $h^2_{t,i}=\mathbb{V}(r_{t,i}|\mathcal{I}_{t-1})$.

Since our approach is developed in a fully semi-parametric framework, the specification for $h^2_{t,i}$ is indirectly recovered assuming an ES-CAViaR type model for the individual asset VaRs and ESs.
Among the several diverse specifications that have been proposed in the literature \citep{engle2004caviar,tayl2019}, for presentation purposes, without any loss of generality, we focus on the ES-CAViaR model with the Indirect GARCH (ES-CAViaR-IG) specification for VaR and the multiplicative VaR to ES relationship:
\begin{eqnarray}\label{e:es-caviar-ig}
   Q_{t,i} &=&- \sqrt{\omega^{(q)}_i +\alpha^{(q)}_i r_{t-1,i}^2+\beta_i  Q^2_{t-1,i}},\\
  ES^2_{t,i}&=&(1+\text{exp}(\gamma_{0,i}))Q^2_{t,i},
    \qquad   i=1,\ldots,n, \nonumber
    \end{eqnarray}
where, by a variance targeting argument, the intercept $\omega^{(q)}_i$ can be parameterized as:
    \begin{equation}
        \omega^{(q)}_i=(q^2_i(1-\beta_i)-\alpha^{(q)}_i)\,\mathbb{V}(r_{t,i}),
        \label{e:vartarget}
    \end{equation}
    and $q_i $ is the VaR factor for the $i$-th asset.
In order to derive (\ref{e:vartarget}), let $Q^2_{t,i}= q^2_{i} h^2_{t,i}$; a standard univariate GARCH implies
\begin{equation}
    h^2_{t,i} =\omega_i +\alpha_i r_{t-1,i}^2+\beta_i  h^2_{t-1,i},\\
    \label{e:vartarget_derive_std}
\end{equation}
where $\omega_i = \omega^{(q)}_i/q^2_{i}$ and $\alpha_i =\alpha^{(q)}_i/q^2_{i}$.
Introducing variance targeting in (\ref{e:vartarget_derive_std}) produces
\begin{equation}
    h^2_{t,i} = (1-\alpha_i-\beta_i)\mathbb{V}(r_{t,i})+\alpha_i r_{t-1,i}^2+\beta_i  h^2_{t-1,i}.\\
    \label{e:vartarget_derive}
\end{equation}
Multiplying both sides of the recursion (\ref{e:vartarget_derive}) by $q^2_{i}$, we have
\begin{equation}
    Q^2_{t,i} = (q^2_i(1-\beta_i)-\alpha^{(q)}_i)\,\mathbb{V}(r_{t,i})+\alpha_i^{(q)} r_{t-1,i}^2+\beta_i  Q^2_{t-1,i}.
    \label{e:vartarget_derive_1}
\end{equation}
Comparing (\ref{e:es-caviar-ig}) and (\ref{e:vartarget_derive_1}), Equation (\ref{e:vartarget}) is obtained.

The next step is to define the dynamic model for $P_t$ which is the conditional correlation matrix of the returns vector $\mathbf{r}_t$. As in \cite{engle2002dynamic}, to model $P_t$ in Equation (\ref{e:dcc1}) and ensure unit correlations on its main diagonal, we adopt the specification:
\begin{equation}\label{e:dcc_pt}
P_t=S^{-1/2}_t R_t S^{-1/2}_t,
\end{equation}
where $S^{1/2}_t$ is a diagonal matrix containing the diagonal entries of $R^{1/2}_t$, i.e., $S^{1/2}_{t,ii}=R^{1/2}_{t,ii}$. The dynamics of $R_t$ can be parsimoniously modelled as:
\begin{equation}
R_t=\Omega+a \boldsymbol{\epsilon}_t\boldsymbol{\epsilon}^{'}_t+ b R_{t-1},
    \label{e:qdcc}
\end{equation}
where $\boldsymbol{\epsilon}_t$ is a ($n \times 1$) vector whose $i$-th element is given by ${\epsilon}_{t,i}=r_{t,i}/h_{t,i}$; $a$ and $b$ are non-negative coefficients satisfying the stationarity condition $(a+b)<1$. When $\Omega$ and $R_0$ are  positive definite and symmetric (PDS), the condition $(a+b)<1$ is sufficient to ensure that  $R_t$ is PDS, for any time point $t$. As further explained in Section \ref{estimation_section}, the maximum number of simultaneously estimated coefficients can be further reduced by applying correlation targeting \citep{engle2002dynamic} in Equation (\ref{e:qdcc}):
\begin{eqnarray}\label{e:dcc_q_targeting}
R_t=(1-a-b)\hat{\Sigma}_{\boldsymbol{\epsilon}}+a \boldsymbol{\epsilon}_{t-1}\boldsymbol{\epsilon}^{'}_{t-1}+b R_{t-1},
\end{eqnarray}
where
\begin{equation*}
\hat{\Sigma}_{\boldsymbol{\epsilon}}=T^{-1}\sum_{t=1}^{T}\boldsymbol{\epsilon}_t\boldsymbol{\epsilon}^{'}_t.
\end{equation*}

Finally, the model is completed by the portfolio VaR and ES specifications
\begin{eqnarray} \label{e:modvar_es_dcc}
Q_{t}=\mathbf{w}^{'}\boldsymbol{\mu}_t +q\sqrt{\mathbf{w}^{'}{D_t P_t D_t}\mathbf{w}},  \qquad
\text{ES}_{t}=\mathbf{w}^{'}\boldsymbol{\mu}_t + c\sqrt{\mathbf{w}^{'}{D_t P_t D_t}\mathbf{w}},
\end{eqnarray}
where we set
\begin{equation}
c^2={q^2\left(1+\exp{(\gamma_0})\right)}.
\label{e:multscal}
\end{equation}

Given the assumptions on $a$ and $b$, no identifiability issue arises and only four coefficients need to be estimated in the second stage of the Semi-DCC model: $q$, $a$, $b$ and $\gamma_0$. The parameter estimate $q$ (VaR factor) allows us to directly estimate the quantile of the error distribution in the model for portfolio returns. Since we have defined $ c^2/q^2 = 1+ \text{exp}(\gamma_0)$, after estimating parameters $q$ and $\gamma_0$, the value of the ES factor $c$ can be immediately recovered. The VaR and ES joint loss function from the FZ class can be incorporated into the proposed Semi-DCC for model estimation. In this paper, we focus on testing AL and FZ0 losses, with the corresponding namings of Semi-DCC-AL and Semi-DCC-FZ0 used.\\

\noindent
\emph{Remark 1.} The proposed model can be further explored with a time varying $c$ to $q$ relationship. However, the results in \cite{tayl2019} and \cite{gerlach2020semi} show that a constant multiplicative factor between VaR and ES is capable of producing very competitive risk forecasts in univariate risk forecasting models. Therefore, to limit the focus of the paper we employ the constant multiplicative factor $(1+\text{exp}(\gamma_0))$.

\section{Estimation}\label{estimation_section}

The estimation of the Semi-DCC model can be performed by means of a two step procedure in the spirit of \cite{engle2002dynamic}. For demonstration purpose, the Semi-DCC using the AL loss is presented in this section, while other VaR and ES joint loss functions such as FZ0 can be directly incorporated, by replacing AL with FZ0 as the optimization target.

\noindent
 \textbf{Step 1} is dedicated to the estimation of the individual assets volatilities. In order to gain robustness against heavy tailed distributions,  the $h_{t,i}$ values are indirectly obtained via the estimation of $n$ separate semi-parametric ES-CAViaR-IG models employing a multiplicative VaR to ES factor, as defined in Equation (\ref{e:es-caviar-ig})\footnote{Alternatively, the $h_{t,i}$ could be obtained by Gaussian QML estimation of GARCH(1,1) models. To maintain the consistency, for both steps 1 and step 2 we use the AL loss based estimation.}.


 Namely, for each asset, the coefficients of the individual risk models \[ \boldsymbol{\xi}_i=( \alpha^{(q)}_i,\beta_i, q_i, \gamma_{0,i} )\]
 are separately estimated  minimizing an AL loss:
    \begin{eqnarray*}
\label{e:dcc_al_stage1}
\widehat{\boldsymbol{\xi}}_i=\underset{\boldsymbol{\xi}_i}{\arg\min}\quad \ell_1(r_i;\boldsymbol{\xi}_i)= - \sum_{t=1}^{T} \left(\text{log}  \frac{(\alpha-1)}{\text{ES}_{t,i}} + {\frac{(r_{t,i}-Q_{t,i})(\alpha-I(r_{t,i}\leq Q_{t,i}))}{\alpha \text{ES}_{t,i}}} \right),
\end{eqnarray*}
for $i=1,\ldots, n$. The set of estimated 1-st stage coefficients is denoted as: $\widehat{\boldsymbol{\xi}}=( \widehat{\boldsymbol{\xi}}_1,\ldots, \widehat{\boldsymbol{\xi}}_n )$.\\

\noindent
\textbf{Step 2} is dedicated to the estimation, conditional on $\widehat{\boldsymbol{\xi}}$, of the coefficients controlling correlation dynamics $( \text{vech}(\Omega),a,b )$ and the tail properties of the conditional distribution of portfolio returns ($q,\gamma_0$). Here, as usual, the notation $\text{vech}(\Omega)$ denotes the column-stacking operator applied to the upper portion of the symmetric matrix $\Omega$ and is used with the only purpose of indicating the unique elements of the matrix $\Omega$ to be estimated.

Again, these coefficients can be jointly estimated minimizing an AL loss function specified in terms of portfolio returns $r_t$:
\begin{eqnarray}
\label{e:dcc_al_stage2}
\widehat{\boldsymbol{\theta}}=\underset{\boldsymbol{\theta}}{\arg\min}\qquad 
\ell_2(r;\boldsymbol{\theta}|\widehat{\boldsymbol{\xi}})= -\sum_{t=1}^{T} \left( \text{log}  \frac{(\alpha-1)}{\text{ES}_t} + {\frac{(r_t-Q_t)(\alpha-I(r_t\leq Q_t))}{\alpha \text{ES}_t}} \right),
\end{eqnarray}
where $Q_t$ and $\text{ES}_t$ are the portfolio VaR and ES as defined in Equation (\ref{e:modvar_es_dcc}), $\boldsymbol{\theta}=$ $(\text{vech}(\Omega),$ $a,b,q,\gamma_0)$.

The estimated volatilities from Step 1 are used to compute the estimated standardized asset returns:
\[
\widehat{\boldsymbol{{\epsilon}}}_t=\widehat{D}^{-1}_t\mathbf{r}_t,
\]
where $\widehat{D}_{t,ii}=\widehat{h}_{t,i}=\widehat{Q}_{t,i}/\hat{q}_i$. $\widehat{\epsilon}_t$ is then plugged into Equation (\ref{e:qdcc}) to obtain $\widehat{R}_t$ that can in turn be used to obtain $\widehat{P}_t$ using Equation (\ref{e:dcc_pt});  Finally, $\hat{q}$, $\hat{c}$, $\widehat{P}_t$ and $\widehat{D}_t$ are used to produce portfolio VaR and ES, $\widehat{Q}_t$ and $\widehat{\text{ES}}_t$, according to Equation (\ref{e:modvar_es_dcc}).\\

\noindent
\emph{Remark 2}. It is worth noting that direct estimation of $\Omega$ would imply optimizing the AL loss wrt $n(n+1)/2$ additional parameters, that can be hardly feasible for even moderately large cross-sectional dimensions. For example, in our empirical analysis, where we work with a portfolio of dimension $n=28$, we would have to estimate 406 distinct elements of $\Omega$. To overcome this issue, correlation targeting can be applied. When correlation targeting is used and $R_t$ is modelled as in Equation (\ref{e:dcc_q_targeting}), the estimation procedure is modified to incorporate an intermediate step in which $\widehat{\Sigma}_{\boldsymbol{{\epsilon}}}$ is estimated by the sample variance and covariance matrix of $\widehat{\boldsymbol{{\epsilon}}}_t=\widehat{D}^{-1}_t\mathbf{r}_t$. Therefore, the loss in Equation (\ref{e:dcc_al_stage2}) is then optimized only wrt four parameters $(a,b,q,\gamma_0)$.\\

\noindent
In step 1, we use the Matlab 2023a ``fmincon'' optimization routine to estimate the ES-CAViaR-IG models.
In step 2, the Matlab ``MultiStart'' facility is further employed for ``fmincon'' to improve the robustness of the estimation results. We use 5 separate sets of starting points, generating 5 local solutions, then the optimum set among these is finally chosen as the parameter estimates. The sensitivity of strictly consistent loss functions, such as AL and FZ0, to initial conditions is a well-known problem that typically complicates the estimation of model parameters in semi-parametric risk models \citep{engle2004caviar}. Nevertheless, estimation strategies based on the use of multiple initial values, such as the one implemented in this paper, have been found in the literature to be effective in overcoming this sensitivity \citep[see][among others]{tayl2019}.

\section{Simulation study}\label{simulation_section}

A simulation study has been conducted in order to assess the statistical properties of the two-step VaR and ES joint loss based estimation procedure discussed in Section \ref{estimation_section}. We focus on presenting the simulation results of Semi-DCC-AL when the AL based loss is used, since we find that the performances of Semi-DCC-AL and Semi-DCC-FZ0 are very close. In the following empirical section, risk forecasting results of both Semi-DCC-AL and Semi-DCC-FZ0 will be presented.

Our simulation analysis has two main objectives. First, the study evaluates the bias and efficiency of the estimators of the Semi-DCC-AL parameters ($a,b,q,\gamma_0$). Second, we assess the the risk forecasting performances of the estimated Semi-DCC-AL model and its comparison to parametric DCC models in an equally weighted portfolio, via comparing the one-step-ahead level portfolio VaR and ES forecast accuracy to the ``true" simulated values.

\subsection{Semi-DCC-AL performance}

In line with the empirical findings arising from our empirical application, we have considered the Data Generating Process (DGP) as a parametric DCC model of dimension $n$=28 with the true values of the correlation dynamic parameters in Equation (\ref{e:dcc_q_targeting}) given by $a=0.12$ and $b=0.78$:
\begin{eqnarray}\label{e:dcc_q_targeting_simu}
	Q_t=(1-0.12-0.78) \Sigma_{\boldsymbol{\epsilon}}+ 0.12 \boldsymbol{\epsilon}_{t-1}\boldsymbol{\epsilon}^{'}_{t-1}+0.78 Q_{t-1},
\end{eqnarray}
where $\Sigma_{\boldsymbol{\epsilon}}$ is a $n \times n$ matrix with unit diagonal and off-diagonal elements equal to 0.5.

In the return equation $\mathbf{r}_t= \boldsymbol{\mu}_t+H^{1/2}_t \mathbf{z}_t$, the conditional mean vector $\boldsymbol{\mu}_t$ of returns is chosen to be zero. The DGP univariate volatilities are assumed to follow the GARCH(1,1) model:
\[
h^2_{t,i}=0.1+0.1r^2_{t-1,i}+0.8 h^2_{t-1,i},\qquad i=1,\ldots,n.
\]
The parametric distribution of returns is assumed to be $\mathbf{z}_t \overset{iid}{\sim} D_1(\mathbf{0},I_n)$, where the following choices of $D_1$ have been considered:
\begin{itemize}
	\item Standardized multivariate Normal distribution: $\mathbf{z}_t \overset{iid}{\sim} \mathcal{N}_n(\mathbf{0},I_n)$.
	\item  Standardized multivariate Student's $t$ distribution: $\mathbf{z}_t\overset{iid}{\sim} t_n(\mathbf{0},I_n;\nu)$, where the degrees of freedom parameter $\nu$ has been set equal to 10.
	\item A multivariate non-spherical distribution with Student's $t$ marginals ($nst$: $\mathbf{z}_t\overset{iid}{\sim} nst_n(\mathbf{0},I_n;\boldsymbol{\nu})$; the density of this distribution is obtained taking the product of $n$ independent univariate standardized $t_1(0,1;{\nu}_i)$ densities; $\boldsymbol{\nu}=(\nu_1,\ldots,\nu_n)$ is the $n$-dimensional vector of marginal degrees of freedom parameters. When $\nu_i=\nu$ ($\forall i$), the marginal densities of the product are the same as those of the multivariate $t$, although the joint density is different \citep{baulaur2005}. In our simulations, the values of $\nu_i$ have been uniformly drawn over the interval [5,15].
\end{itemize}
While the first two distributions belong to the spherical family, the same does not hold for the third one ($nst$).

On the $\alpha=2.5\%$ probability level\footnote{We have also tested the $1\%$ and $5\%$ probability levels, while the observations on the Semi-DCC-AL parameter estimates and risk forecasts are in general consistent with the ones from $2.5\%$, thus only the 2.5\% results are presented in this section. In the following section, we compare the risk forecasting performance of the standard parametric DCC models with the Semi-DCC-AL on the $1\%$, $2.5\%$ and $5\%$ levels in detail.}, the Semi-DCC-AL model is then fitted to the time series of equally weighted $(\mathbf{w}=1/n)$ portfolio returns generated from the chosen DGP, for three different sample sizes $T\in\{2000,3000,5000\}$. Overall, matching 3 different distributional assumptions (including both spherical and non-spherical distributions) and sample sizes, 9 simulation settings have been obtained and, under each of these, 250 return series have been generated.

Conditional on the chosen error distribution and on the values of the correlation parameters $a$ and $b$, the true values of the other two parameters $q$ and $\gamma_0$ in the Semi-DCC-AL model can be then calculated as follows.

In the spherical case, multivariate Normal and $t$, letting $F(.)$ be the CDF of $z_{t,i}$, $q=F^{-1}(\alpha)$ and $c=E(z_{t,i}|z_{t,i} \le q)$. By Equation (\ref{e:multscal}) it is then easy to obtain:
\[
\gamma_0=\log\left(\frac{c^2}{q^2}-1\right).
\]

Therefore, taking the multivariate Normal case on the $\alpha=2.5\%$ probability as an example, in the Semi-DCC-AL the true value of $q^2$ is $\Phi^{-1}(\alpha)^2= (-1.96)^2= 3.8415$. The true value for  $c^2$ is $\left( \frac{\phi(\Phi^{-1}(\alpha))}{\alpha} \right)^2= (-2.3378)^2$, where $\Phi^{}$ is the standard Normal CDF and $\phi$ is standard Normal Probability Density Function (PDF). Further, we have $ (1+ \text{exp}(\gamma_0)) = \frac{c^2}{q^2} = \frac{(-2.3378)^2}{ (-1.9600)^2}= 1.4227$, thus $\gamma_0= \text{log} (1.4227-1)= -0.8610$. These true values of $a$, $b$, $\gamma_0$ and $q^2$ for the multivariate Normal distribution are shown in the \emph{True} rows in Table \ref{simu_table_dcc}. When $\mathbf{z}_t\overset{iid}{\sim} t_n(\mathbf{0},I_n;\nu=10)$, the true values of $q$, $c$ and $\gamma_0$ are obtained through a similar approach, replacing the analytical formulas for $q$ and $c$ with the equivalent expressions for an univariate $t$ distribution.

In the $nst$ case, the true values of $q$, $c$ and $\gamma_0$ need to be calculated via simulations since this distribution is not a member of the spherical family. It also follows that, differently from what observed for the multivariate Normal and $t$ cases, the values of these coefficients will be dependent on the chosen portfolio allocation, thus they could change when moving from equal weighting to a different allocation scheme.

More specifically, $q$ and $c$ can be calculated by the following procedure.

\begin{enumerate}
	\item For a given set of randomly generated degrees of freedom values $\boldsymbol{\nu}$ and portfolio allocation $\mathbf{w}$, a return series with size $T_{sim}\times n$ ($T_{sim}=10^5$) is simulated from the specific DCC process taken as DGP: $\mathbf{r}^{*}_{j}$, $j=1,\ldots,T_{sim}$.
	\item Given the simulated returns vector $\mathbf{r}^{*}_{j}$, conditional covariance matrix ${H}^{*}_j$ and portfolio allocation $\mathbf{w}$, the time series of simulated standardized portfolio returns is computed as:
	\[
	z^*_j=\frac{r^*_j}{h^*_{j}},
	\]
	for $j=1,\ldots,T_{sim}$ and where
	\[
	r^*_j=\mathbf{w}' \mathbf{r^*_j}, \qquad h^*_{j}=\sqrt{\mathbf{w}'{H}^{*}_j\mathbf{w}}.
	\]
	\item The empirical quantile and conditional tail average of the $  z^*_j$ series are then  used as the simulated ``true'' values for $q$ and $c$. These values can be used to calculate the true values for $\gamma_0$ through inverting Equation (\ref{e:multscal}).
	%
\end{enumerate}
For ease of reference, for each DGP, the true parameter values for $(a,b,q,\gamma_0)$ are included in the \emph{True} rows in Table \ref{simu_table_dcc}.

The Semi-DCC-AL model is then fitted to each simulated dataset, using the estimation procedure as outlined in Section \ref{estimation_section}. The simulation results for the Semi-DCC-AL model are shown in Table \ref{simu_table_dcc} where, for ease of presentation, we have chosen not to report parameter estimates of the fitted first stage ES-CAViaR-IG models.

The rows labeled as \emph{True} report, for each DGP, the values of the parameter true values used for simulation. The empirical averages of the 250 parameter estimates, for various return distributions and sample sizes, are shown in the \emph{Mean} rows. The Root Mean Squared Error (RMSE) values between the parameter estimates and the true values are shown in the \emph{RMSE} rows in Table \ref{simu_table_dcc}.

To further evaluate the accuracy of the portfolio VaR and ES forecasts from the estimated Semi-DCC-AL model, for all 250 simulated datasets, we compare the one-step-ahead $\widehat{Q}_{T+1}$ and $\widehat{\text{ES}}_{T+1}$ forecasts based on the \emph{estimated} parameters with their counterparts based on the \emph{true} DGP coefficients. For $\widehat{Q}_{T+1}$ and $\widehat{\text{ES}}_{T+1}$, the \emph{True} rows then report the averages of the 250 risk forecasts based on the \emph{true} parameters, for different return distributions and sample sizes.
First, the results provide support to the use of the two-step AL based estimation method and show that it is able to produce relatively accurate parameter estimates. Overall, the estimated bias  ($|\emph{Mean}-\emph{True}|$) is reasonably low. When the distribution is fixed, the absolute bias decreases as the time series length $T$ increases. For all parameters, the values of the RMSE also monotonically decrease as $T$ increases, suggesting consistency of the estimation procedure.
When the sample size is fixed, as expected, the RMSE values are clearly larger when the errors follow a multivariate Student's $t$, comparing to multivariate Normal distribution.

The estimation results are still relatively accurate for the non-spherical $nst$ distribution that, for correlation parameters, returns RMSE values slightly higher than those obtained in the Normal case. For parameters $q$ and $\gamma_0$, the simulated RMSE is in line with the Normal case, for $\gamma_0$, and even lower for $q$. 

These results suggest that the proposed semi-parametric estimation procedure is able not only to keep track of the volatility and correlation dynamics but also of the distributional properties of portfolio returns, through the estimation of $q$ and $\gamma_0$ (implicitly $c$).

Finally, reminding that the main motivation for the Semi-DCC-AL model is the generation of accurate portfolio risk forecasts, the last two columns of Table \ref{simu_table_dcc} provide a very important benchmark for assessing the properties of the proposed estimation procedure. Comparing the $\alpha=2.5\%$ \emph{estimated} and \emph{true} risk forecasts, for both VaR and ES, it can be noted that these two series are on average very close even for the shortest sample size $T=2000$. The RMSE values are also remarkably low: with $T=2000$ they do not exceed 0.0811 for VaR and 0.1034 for ES, and their values monotonically decrease as $T$ increases across three return distributions. This last set of results confirms the ability of the proposed two-stage estimation procedure to accurately reproduce the portfolio risk dynamics.

\begin{table}[!ht]
	\begin{center}
		\caption{\label{simu_table_dcc} Semi-DCC-AL model parameter estimates and VaR and ES forecasting results with simulated datasets.}
		\tabcolsep=10pt
		\scriptsize
		\vspace{10pt}
		\begin{tabular}{lccccccccc} \hline
			DGP&& $a$ &$b$ & $\gamma_0$ & $q^2$ & $\widehat{Q}_{T+1}$ &$\widehat{\text{ES}}_{T+1}$ \\ \hline
			$\mathcal{N}$, $T=2000$&True&0.1200&0.7800&-0.8610&3.8415&-1.3704&-1.6346\\
			&Mean&0.1360&0.7150&-0.9131&3.9308&-1.3849&-1.6420\\
			&RMSE&0.0751&0.1995&0.1639&0.2057&0.0606&0.0762\\
			$\mathcal{N}$, $T=3000$&True&0.1200&0.7800&-0.8610&3.8415&-1.3565&-1.6180\\
			&Mean&0.1310&0.7319&-0.8919&3.9079&-1.3669&-1.6244\\
			&RMSE&0.0615&0.1736&0.1200&0.1613&0.0521&0.0611\\
			$\mathcal{N}$, $T=5000$&True&0.1200&0.7800&-0.8610&3.8415&-1.3665&-1.6299\\
			&Mean&0.1265&0.7528&-0.8830&3.9018&-1.3760&-1.6367\\
			&RMSE&0.0466&0.1118&0.1029&0.1363&0.0392&0.0454\\ \hline
			$t$, $T=2000$&True&0.1200&0.7800&-0.5097&3.9717&-1.4361&-1.8169\\
			&Mean&0.1591&0.6766&-0.5839&4.1157&-1.4499&-1.8135\\
			&RMSE&0.1033&0.2437&0.1894&0.2954&0.0818&0.1045\\
			$t$, $T=3000$&True&0.1200&0.7800&-0.5097&3.9717&-1.3825&-1.7491\\
			&Mean&0.1497&0.6894&-0.5658&4.0686&-1.3990&-1.7543\\
			&RMSE&0.0884&0.2248&0.1527&0.2167&0.0692&0.0850\\
			$t$, $T=5000$&True&0.1200&0.7800&-0.5097&3.9717&-1.4155&-1.7909\\
			&Mean&0.1303&0.7395&-0.5459&4.0619&-1.4304&-1.7995\\
			&RMSE&0.0617&0.1564&0.1172&0.1613&0.0534&0.0703\\ \hline
			$nst$, $T=2000$&True&0.1200&0.7800&-0.8311&3.8772&-1.3570&-1.6259\\
			&Mean&0.1440&0.7030&-0.8893&3.9012&-1.3538&-1.6096\\
			&RMSE&0.0894&0.2245&0.1656&0.1849&0.0639&0.0762\\
			$nst$, $T=3000$&True&0.1200&0.7800&-0.8311&3.8772&-1.3806&-1.6542\\
			&Mean&0.1356&0.7223&-0.8657&3.8978&-1.3799&-1.6461\\
			&RMSE&0.0729&0.1926&0.1246&0.1539&0.0576&0.0706\\
			$nst$, $T=5000$&True&0.1200&0.7800&-0.8311&3.8772&-1.4001&-1.6775\\
			&Mean&0.1277&0.7486&-0.8569&3.9091&-1.4057&-1.6788\\
			&RMSE&0.0496&0.1195&0.1029&0.1251&0.0383&0.0465\\
			
			\hline
		\end{tabular}
	\end{center}
\end{table}

Summarizing, the analysis of simulation results reveals some encouraging regularities:
\begin{itemize}
    \item[-] the absolute value of the estimated bias is, for all coefficients, small (in relative terms);
    \item[-] the simulated Root Mean Squared Error (RMSE) monotonically decreases with the sample size $T$, suggesting consistency of the estimation procedure;
    \item[-] as expected, the RMSE values tend to increase when heavier tailed distributions are considered;
    \item[-] the predicted VaR and ES are on average very close to their simulated counterparts.
\end{itemize}

\subsection{Standard DCC and Semi-DCC risk forecasting comparison}

As mentioned, generating accurate portfolio risk forecasts via a semi-parametric approach is a key motivation when proposing the Semi-DCC framework. Therefore, to further demonstrate the effectiveness of the Semi-DCC in risk forecasting with the AL loss, the performance of Semi-DCC-AL is compared to the standard parametric DCC models, on the $1\%$, $2.5\%$ and $5\%$ probability levels. We first use the simulated datasets following the standardized multivariate Student's $t$ distribution as in the previous section $\mathbf{z}_t\overset{iid}{\sim} t_n(\mathbf{0},I_n;\nu)$ with the degrees of freedom parameter $\nu=10$. Then a heavier tailed choice of $\nu=5$ is also included. The Student's $t$ is selected, as it can control the tail of the returns and the simulation results shown in the previous section with Normal and $nst$ are in general consistent with the ones from Student's $t$.

The VaR and ES forecasting performance of the proposed Semi-DCC-AL framework on the 250 simulated datasets are compared to those yielded by the conventional DCC models fitted through 18 different DGPs which include: three probability levels $1\%$, $2.5\%$ and $5\%$ ; three different sample size $T\in\{2000,3000,5000\}$; two degrees of freedom parameters $\nu\in\{5,10\}$. For comparison, first we consider a standard parametric DCC-N and DCC-t models fitted by a two-stage procedure \citep{engle2002dynamic} combining maximization of Normal and Student's $t$ distributions likelihoods in the first (volatility) and second (correlation) stages of the estimation procedure with the application of correlation targeting. For the VaR and ES forecasting, the theoretical quantile and tail expectation based on the Normal or Student's $t$ distributions are used as detailed in the previous section. Further, the Composite Likelihood (clik) approach developed in \cite{pakel2021fitting} using Normal distribution is also included for comparison (DCC-clik-N).

Following the VaR and ES comparison setup used in the previous section, we report the \emph{true} $\widehat{Q}_{T+1}$ and $\widehat{\text{ES}}_{T+1}$ forecasts as the averages of the 250 risk forecasts, based on the \emph{true} parameters for different return distributions and sample sizes. Then the Mean and RMSE values of VaR and ES forecasts from each model under each DGP are calculated to illustrate the risk forecasting performance of different models. The VaR and ES forecasting results are presented in Tables \ref{simu_table_var} and \ref{simu_table_es} respectively. Overall, based on various DGPs the ES forecasting bias and RMSE results of the proposed Semi-DCC-AL clearly outperform those of the parametric DCC models. For VaR forecasting, regarding bias ($|\emph{Mean}-\emph{True}|$) the Semi-DCC-AL in general outperform the standard DCC models, and for RMSE the Semi-DCC-AL is preferred for the more extreme 1\% level and $\nu=5$. Even if the DGPs follow the Student's $t$ distribution, the risk forecasts, in particular ES, from the Semi-DCC-AL still generate less bias and smaller RMSE values than the ones from DCC-t. Such observation demonstrates the effectiveness of the proposed semi-parametric approach in producing accurate portfolio risk forecasting. Other observations are summarized below:
\begin{itemize}
    \item[-] Across different probability levels, it can be seen that the more extreme 1\% level creates more challenges for the parametric models, e.g., with $T=3000$, $\nu=5$, $1\%$ the ES forecasting RMSE values from parametric DCC models are more than doubled of the one from Semi-DCC-AL. On the least extreme 5\% level, the performance gap of between parametric DCC and Semi-DCC-AL is smaller. Such findings confirm that the proposed semi-parametric approach is more robust on the more extreme probability level.
    \item[-] Across the two different degrees of freedoms, the heavier tail choice of $\nu=5$ poses more challenges for all models, especially for the parametric DCC models. With $T=3000$ and $5\%$, when $\nu=10$ goes to $\nu=5$ the RMSE of ES forecast from parametric DCC models rises from $\approx 0.06$ to $\approx0.11$ for parametric DCC models, while for Semi-DCC-AL the figure goes from 0.0579 to 0.0755. This demonstrates the usefulness of the semi-parametric approach in risk forecasting on the heavy tailed distributions, which are commonly observed in real financial time series data.
    \item[-] Across different sample sizes, consistent with the study in the previous section, the RMSE of Semi-DCC-AL risk forecasts decreases as the time series length $T$ increases.
\end{itemize}

To conclude, the results of the simulation study support the two-step AL based estimation method which is able to produce accurate parameter estimates and portfolio risk forecasts, for both spherical and non-spherical error distributions. Comparing the risk forecasting results of the Semi-DCC-AL to several benchmarking parametric DCC models, the advantage of the proposed semi-parametric approach is clearly demonstrated, especially for the heavier-tailed return distribution and more extreme probably level.

\begin{table}[!ht]
	\begin{center}
		\caption{\label{simu_table_var} VaR forecasting results of parametric DCC models and Semi-DCC-AL with 250 simulation datasets. \cred{Red} highlights RMSE values that are higher than the one from Semi-DCC-AL for each DGP.}
		\tabcolsep=10pt
		\scriptsize
		\vspace{10pt}
		\begin{tabular}{lccccccccc} \hline

DGP&&True-VaR&DCC-N&DCC-t&DCC-clik-N&Semi-DCC-AL\\ \hline
$t$, $T=2000$, $\nu=5$, $1\%$&Mean&-1.7663&-1.5961&-1.6502&-1.5953&-1.7978\\
&RMSE&&\cred{0.1803}&0.1537&\cred{0.1802}&0.1802\\
$t$, $T=2000$, $\nu=5$, $2.5\%$&Mean&-1.3545&-1.3450&-1.3669&-1.3443&-1.3792\\
&RMSE&&0.0285&0.0616&0.0282&0.1067\\
$t$, $T=2000$, $\nu=5$, $5\%$&Mean&-1.0609&-1.1268&-1.1315&-1.1262&-1.0797\\
&RMSE&&\cred{0.0693}&\cred{0.0812}&\cred{0.0692}&0.0691\\
$t$, $T=2000$, $\nu=10$, $1\%$&Mean&-1.7766&-1.6839&-1.7215&-1.684&-1.7984\\
&RMSE&&0.1023&0.0998&0.1016&0.1374\\
$t$, $T=2000$, $\nu=10$, $2.5\%$&Mean&-1.4361&-1.4180&-1.4253&-1.4181&-1.4499\\
&RMSE&&0.0292&0.0611&0.0287&0.0818\\
$t$, $T=2000$, $\nu=10$, $5\%$&Mean&-1.1699&-1.1904&-1.1818&-1.1905&-1.1775\\
&RMSE&&0.0252&0.0464&0.0255&0.0552\\ \hline
$t$, $T=3000$, $\nu=5$, $1\%$&Mean&-1.7729&-1.6007&-1.6522&-1.6003&-1.8002\\
&RMSE&&\cred{0.1784}&\cred{0.1474}&\cred{0.1786}&0.1342\\
$t$, $T=3000$, $\nu=5$, $2.5\%$&Mean&-1.3615&-1.3486&-1.3682&-1.3482&-1.3832\\
&RMSE&&0.0272&0.0537&0.0279&0.0775\\
$t$, $T=3000$, $\nu=5$, $5\%$&Mean&-1.0682&-1.1318&-1.1341&-1.1315&-1.0881\\
&RMSE&&\cred{0.0669}&\cred{0.0757}&\cred{0.0671}&0.0548\\
$t$, $T=3000$, $\nu=10$, $1\%$&Mean&-1.7108&-1.6183&-1.6686&-1.6176&-1.7322\\
&RMSE&&0.0970&0.0834&0.0974&0.1056\\
$t$, $T=3000$, $\nu=10$, $2.5\%$&Mean&-1.3825&-1.3634&-1.3817&-1.3629&-1.3990\\
&RMSE&&0.0252&0.0534&0.0256&0.0692\\
$t$, $T=3000$, $\nu=10$, $5\%$&Mean&-1.1257&-1.1442&-1.1453&-1.1437&-1.1368\\
&RMSE&&0.0218&0.0454&0.0219&0.0469\\ \hline
$t$, $T=5000$, $\nu=5$, $1\%$&Mean&-1.7736&-1.5985&-1.6544&-1.5977&-1.8067\\
&RMSE&&\cred{0.1857}&\cred{0.1659}&\cred{0.1860}&0.1261\\
$t$, $T=5000$, $\nu=5$, $2.5\%$&Mean&-1.3583&-1.3467&-1.3699&-1.3461&-1.3779\\
&RMSE&&0.0266&0.0649&0.0269&0.0724\\
$t$, $T=5000$, $\nu=5$, $5\%$&Mean&-1.0658&-1.1302&-1.1355&-1.1297&-1.0825\\
&RMSE&&0.0697&0.0808&0.0695&0.0452\\
$t$, $T=5000$, $\nu=10$, $1\%$&Mean&-1.7535&-1.6564&-1.6977&-1.6560&-1.7760\\
&RMSE&&0.1008&0.0907&0.1011&0.0892\\
$t$, $T=5000$, $\nu=10$, $2.5\%$&Mean&-1.4155&-1.3955&-1.4058&-1.3952&-1.4304\\
&RMSE&&0.0232&0.0525&0.0235&0.0534\\
$t$, $T=5000$, $\nu=10$, $5\%$&Mean&-1.1515&-1.1712&-1.1652&-1.1709&-1.1652\\
&RMSE&&0.0214&\cred{0.0415}&0.0215&0.0387\\	
			\hline
		\end{tabular}
	\end{center}
\end{table}

\begin{table}[!ht]
	\begin{center}
		\caption{\label{simu_table_es} ES forecasting results of parametric DCC models and Semi-DCC-AL with 250 simulation datasets. \cred{Red} highlights RMSE values that are higher than the one from Semi-DCC-AL for each DGP.}
		\tabcolsep=10pt
		\scriptsize
		\vspace{10pt}
		\begin{tabular}{lccccccccc} \hline
DGP&&True-ES&DCC-N&DCC-t&DCC-clik-N&Semi-DCC-AL\\ \hline
$t$, $T=2000$, $\nu=5$, $1\%$&Mean&-2.3372&-1.8286&-1.9336&-1.8277&-2.3097\\
&RMSE&&\cred{0.5214}&\cred{0.4337}&\cred{0.5213}&0.2613\\
$t$, $T=2000$, $\nu=5$, $2.5\%$&Mean&-1.8555&-1.6043&-1.6672&-1.6035&-1.8663\\
&RMSE&&\cred{0.2578}&\cred{0.2138}&\cred{0.2578}&0.1457\\
$t$, $T=2000$, $\nu=5$, $5\%$&Mean&-1.5217&-1.4131&-1.4506&-1.4123&-1.5368\\
&RMSE&&\cred{0.1136}&\cred{0.1005}&\cred{0.1138}&0.0963\\
$t$, $T=2000$, $\nu=10$, $1\%$&Mean&-2.1620&-1.9291&-2.0173&-1.9293&-2.1411\\
&RMSE&&\cred{0.2412}&\cred{0.1812}&\cred{0.2404}&0.1725\\
$t$, $T=2000$, $\nu=10$, $2.5\%$&Mean&-1.8169&-1.6913&-1.7384&-1.6915&-1.8135\\
&RMSE&&\cred{0.1310}&\cred{0.1145}&\cred{0.1303}&0.1045\\
$t$, $T=2000$, $\nu=10$, $5\%$&Mean&-1.5546&-1.4929&-1.5153&-1.4930&-1.5558\\
&RMSE&&0.0659&\cred{0.0766}&0.0653&0.0733\\ \hline
$t$, $T=3000$, $\nu=5$, $1\%$&Mean&-2.3459&-1.8339&-1.9362&-1.8334&-2.3275\\
&RMSE&&\cred{0.5203}&\cred{0.4310}&\cred{0.5203}&0.2024\\
$t$, $T=3000$, $\nu=5$, $2.5\%$&Mean&-1.8652&-1.6086&-1.6689&-1.6081&-1.8778\\
&RMSE&&\cred{0.2616}&\cred{0.2161}&\cred{0.2617}&0.1146\\
$t$, $T=3000$, $\nu=5$, $5\%$&Mean&-1.5320&-1.4193&-1.4542&-1.4189&-1.5493\\
&RMSE&&\cred{0.1164}&\cred{0.1016}&\cred{0.1165}&0.0755\\
$t$, $T=3000$, $\nu=10$, $1\%$&Mean&-2.0819&-1.8540&-1.9555&-1.8532&-2.0688\\
&RMSE&&\cred{0.2328}&\cred{0.1583}&\cred{0.2332}&0.1265\\
$t$, $T=3000$, $\nu=10$, $2.5\%$&Mean&-1.7491&-1.6262&-1.6855&-1.6256&-1.7543\\
&RMSE&&\cred{0.1262}&\cred{0.0977}&\cred{0.1265}&0.0850\\
$t$, $T=3000$, $\nu=10$, $5\%$&Mean&-1.4958&-1.4349&-1.4686&-1.4343&-1.5021\\
&RMSE&&\cred{0.0637}&\cred{0.0658}&\cred{0.0641}&0.0579\\ \hline
$t$, $T=5000$, $\nu=5$, $1\%$&Mean&-2.3468&-1.8313&-1.9390&-1.8305&-2.3681\\
&RMSE&&\cred{0.5330}&\cred{0.4513}&\cred{0.5334}&0.1949\\
$t$, $T=5000$, $\nu=5$, $2.5\%$&Mean&-1.8608&-1.6063&-1.6712&-1.6056&-1.8824\\
&RMSE&&\cred{0.2616}&\cred{0.2235}&\cred{0.2620}&0.1105\\
$t$, $T=5000$, $\nu=5$, $5\%$&Mean&-1.5286&-1.4173&-1.4561&-1.4167&-1.5509\\
&RMSE&&\cred{0.1145}&\cred{0.1084}&\cred{0.1149}&0.0671\\
$t$, $T=5000$, $\nu=10$, $1\%$&Mean&-2.1339&-1.8977&-1.9899&-1.8972&-2.1370\\
&RMSE&&\cred{0.2406}&\cred{0.1723}&\cred{0.2409}&0.1202\\
$t$, $T=5000$, $\nu=10$, $2.5\%$&Mean&-1.7909&-1.6645&-1.7150&-1.6642&-1.7995\\
&RMSE&&\cred{0.1286}&\cred{0.1043}&\cred{0.1289}&0.0703\\
$t$, $T=5000$, $\nu=10$, $5\%$&Mean&-1.5301&-1.4687&-1.4942&-1.4684&-1.5415\\
&RMSE&&\cred{0.0628}&\cred{0.0675}&\cred{0.0631}&0.0496\\	
			\hline
		\end{tabular}
	\end{center}
\end{table}

\clearpage
\section{Empirical study}\label{empirical_section}

The performance of the Semi-DCC model in portfolio risk forecasting has been assessed via an empirical study on a multivariate time series of US stock returns. After providing a short description of the data and describing the forecasting design in Section \ref{data_empirical}, we assess the risk forecasting performance of the proposed models for an equally weighted portfolio in Section \ref{univariate_risk_empirical}. This choice does not imply any loss of generality since our investigation could be easily replicated under alternative portfolio configurations. Our preference for the equally weighted scheme is motivated by the robust performance of this simple allocation rule that, in many instances, has been found to be competitive with more sophisticated benchmarks \citep{DeMiguel2007}. In addition, in Appendix \ref{po_empirical} we also evaluate the effectiveness of the proposed framework in generating minimum risk portfolios, a task that cannot be accomplished with univariate risk forecasting models.

\subsection{Equally weighted portfolio risk forecasting}\label{data_empirical}

Daily closing price data are collected for 28 of the 30 components of the Dow Jones index, for the period from 4 January 2005 to 28 September 2021. Only assets providing full coverage of the period of interest have been considered for the analysis.



\begin{figure}[htp]
     \centering
\includegraphics[width=0.8\textwidth]{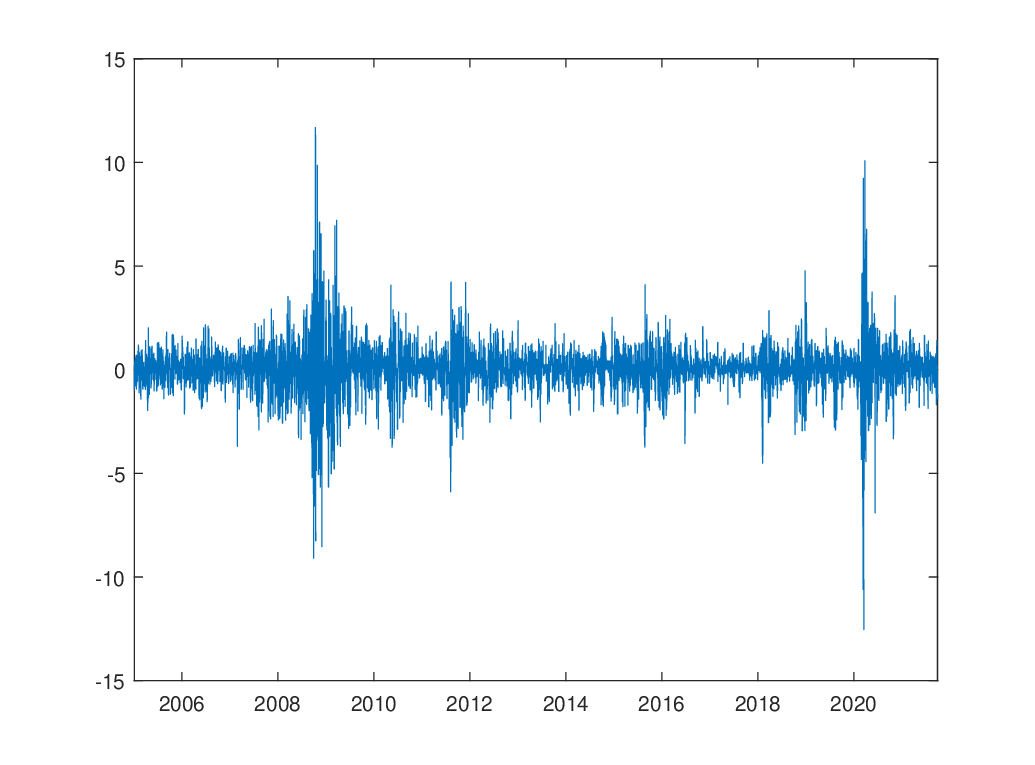}
\caption{\label{f:equal_return} Equally weighted portfolio returns for 28 assets in the Dow Jones index.}
\end{figure}

The computed time series of equally weighted portfolio returns is shown in Figure \ref{f:equal_return}. As can be seen in the plot, there are two major periods of high-volatility. The outbreak of coronavirus disease (COVID-19) has caused a highly volatile period in 2020 and, less recently, the 2008 Global Financial Crisis has also greatly impacted the financial market.

A rolling window scheme, with fixed in-sample size $T_{\text{in}}=3000$ and daily re-estimation, is then implemented to generate $T_{\text{out}}=1213$ out-of-sample one-step-ahead forecasts of VaR and ES at 2.5\% level. Therefore, the in-sample period is from 4 January 2005 to 1 December 2016, and the out-of-sample period covers the time range from 2 December 2016 to 28 September 2021.

Several existing univariate semi-parametric models are selected as benchmarks for comparison. First, we consider the recently proposed ES-CAViaR model \citep{tayl2019} with Symmetric Absolute Value (SAV) and Indirect GARCH (IG) specifications for the quantile regression equation. Regarding the ES specification, to facilitate the performance comparison with the Semi-DCC, we choose the model with the constant multiplicative ES to VaR factor $(1+\text{exp}(\gamma_0) )$, which is the factor that we use in the proposed Semi-DCC model. Two VaR and ES joint loss functions, AL (Semi-DCC-AL) and FZ0 (Semi-DCC-FZ0), are employed and tested.

In addition, the semi-parametric Conditional Autoregressive Expectile framework (CARE, \citealt{tayl2008}), with SAV and IG specifications, is also included.

Furthermore, the risk forecasting performances of the proposed framework are compared to those yielded by the conventional DCC models fitted through different estimation approaches. First, we consider a standard DCC model fitted by a two-stage procedure \citep{engle2002dynamic} combining maximization of Gaussian likelihoods in the first (volatility) and second (correlation) stages of the estimation procedure with the application of correlation targeting. At the VaR and ES forecasting stage, depending on the assumptions formulated on the error distribution, two different series of forecasts are generated and labeled as DCC-N, when a multivariate Normal distribution is assumed, and DCC-QML, when a semi-parametric approach is taken. Namely, for the DCC-N approach, the theoretical quantile and tail expectation based on the Normal distribution are used for VaR and ES calculation. For the QML, differently, the semi-parametric filtered historical simulation approach is used to calculate the VaR and ES. The error quantiles $\hat{q}$ and tail expectations $\hat{c}$ are then estimated by computing the relevant sample quantiles and tail averages of standardized returns ($r_t$ divided by its volatility). Finally, level-$\alpha$ VaR and ES forecasts are obtained by multiplying $\hat{q}$ and $\hat{c}$, respectively, by the portfolio conditional standard deviation forecast from the fitted DCC model.

When applied to even moderately large datasets, such as the one that is here considered, the original approach to the estimation of DCC parameters described in \cite{engle2002dynamic} has been found to be prone to return biased estimates of the correlation dynamic parameters. This motivates our choice to consider, as a further benchmark, the Composite Likelihood (clik) approach developed in \cite{pakel2021fitting}. Along the same lines discussed above, the estimated volatility and correlation parameters are then used to generate two different sets of VaR and ES forecasts labeled as DCC-clik-N and DCC-clik-QML, respectively.

Finally, the picture is completed by considering a parametric DCC model fitted by ML using multivariate Student's $t$ likelihoods in the estimation of volatility and correlation parameters (DCC-t). VaR and ES forecasts are generated considering theoretical quantiles and tail expectations for a standardized Student's $t$ distribution.

For all the DCC benchmarks, in order to guarantee a fair comparison with the Semi-DCC model, the fitted univariate volatility specifications are given by GARCH(1,1) models.

\begin{figure}[htb]
		\begin{center}
			\includegraphics[width=0.8\textwidth]{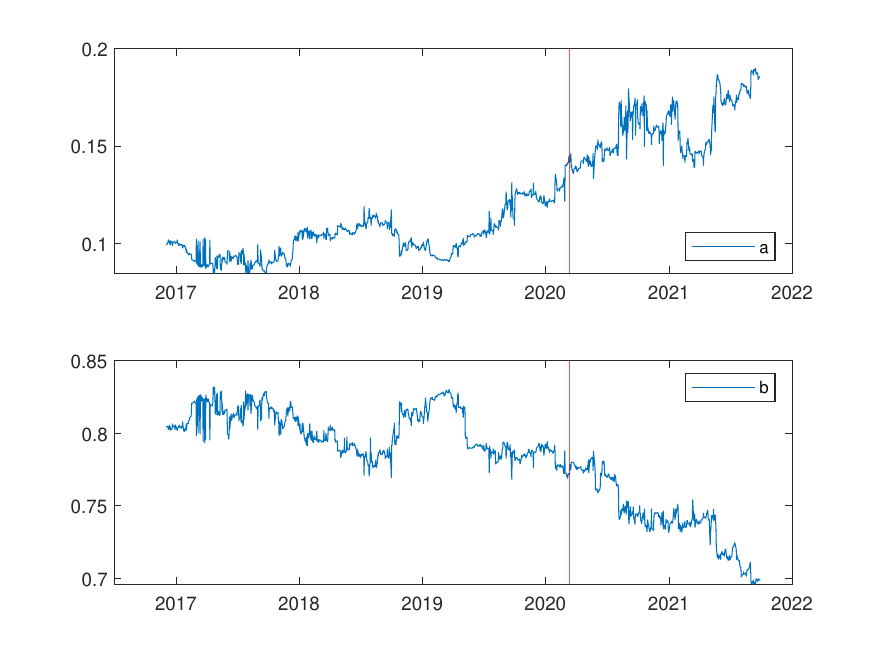}
		\end{center}
		\caption {\label{f:ab_tplots} Plots of estimated correlation parameters $a$ and $b$ of the Semi-DCC-AL model on the 2.5\% probability level. The vertical lines denote the outbreak of the COVID-19 emergency (as officially declared by the World Health Organization on 30 January 2020).
		Whole forecasting sample (1213 days).}		
	\end{figure}

In Figure (\ref{f:ab_tplots}), it is interesting to note that the estimated correlation parameters in the Semi-DCC-AL model vary over the forecasting period, reacting to changes in the underlying market volatility level. In particular, the estimated value of $a$ is characterized by a positive trend originating at the outbreak of the pandemic COVID-19 crisis. An opposite behaviour is observed for $b$.

It is worth noting that the negative correlation emerges as an empirical property of the conditional correlation process. This behaviour is consistent with the recent literature, which provides evidence that conditional correlations have a state-dependent nature, driven mainly by the underlying volatility dynamics \citep[see][among others]{bauwotra2016}. The link between the level of volatility and correlation dynamics is particularly evident in crisis periods, such as the one following the outbreak of the COVID 19 pandemics.

\begin{figure}[htb]
		\begin{center}
			\includegraphics[width=0.7\textwidth]{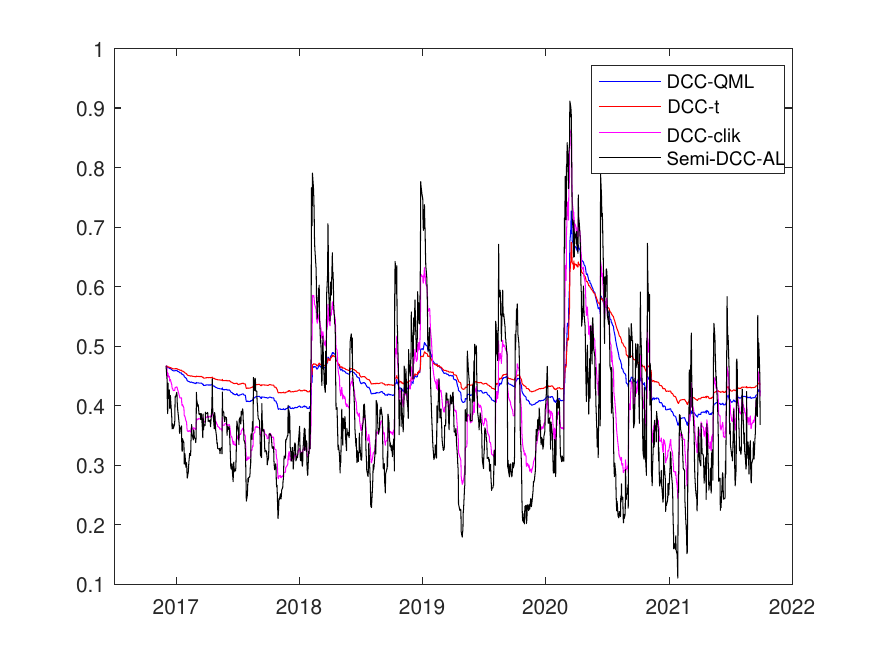}
		\end{center}
		\caption{\label{f:corr_tplots} Cross-sectional averages of the estimated conditional correlation for DCC-QML, DCC-t, DCC-clik and Semi-DCC-AL on the 2.5\% probability level, across the whole forecasting sample (1213 days).}		
\end{figure}

Table \ref{t:corrcoeffs} compares the Semi-DCC-AL estimates (2.5\% probability level) of correlation coefficients with those obtained for the benchmark DCC specifications considered, reporting the average estimates of correlation parameters $a$ and $b$ across all forecasting steps. Here, consistent with previous findings in the literature, standard DCC estimators tend to oversmooth (estimated $b$ very close to 1) correlations, while this is not the case for Semi-DCC-AL models. The DCC-clik estimates stay in between. It is here worth noting that the DCC-clik and Semi-DCC-AL models are based on different first stage volatility estimators: QML-GARCH for DCC-clik, and ES-CAViaR-IG as in Equation (\ref{e:es-caviar-ig}) for Semi-DCC-AL. In addition, Figure \ref{f:corr_tplots} shows the cross-sectional averages of the estimated conditional correlations (calculated based on $P_t$ in Equation (\ref{e:dcc_pt})) across the forecasting period. As can be seen, the dynamics of the estimated conditional correlations between Semi-DCC-AL and DCC-clik are relatively close to each other.

\begin{table}[htb]
	\caption{\label{t:corrcoeffs} Average estimates (across time) of correlation dynamic parameters ($a$ and $b$) from different models and estimation methods on the 2.5\% probability level.  Whole forecasting sample (1213 days).}
	\begin{center}
\begin{tabular}{ccccc}
	& Semi-DCC-AL & DCC-QML & DCC-t &DCC-clik\\
	\hline
$a$ &0.1217   & 0.0040 &   0.0023  &  0.0308\\
$b$ & 0.7827   & 0.9788 &   0.9816  &  0.9198\\
\hline
\end{tabular}
\end{center}
\end{table}

The average of estimated first stage univariate modeling coefficients on the 2.5\% probability level, across all assets and forecasting steps, is reported in Table \ref{t:volcoeffs}. GARCH-$t$ and QML-GARCH, used in DCC-t and DCC-QML respectively, appear to be less reactive (smaller $\alpha$ estimates) to past shocks compared to the ES-CAViaR-IG, used in the Semi-DCC-AL model.

\begin{table}
	\caption{\label{t:volcoeffs} Average estimates (across times and assets) of the first stage univariate modeling parameters ($\alpha$ and $\beta$) from different models and estimation methods on the 2.5\% probability level, across the full forecasting sample (1213 days). For ES-CAVIAR-IG, we report $\alpha=\alpha^{(q)}/q^2$.}
	\begin{center}
		\begin{tabular}{cccc}
			& GARCH-$t$ & QML-GARCH & ES-CAViaR-IG \\
			\hline
		$\alpha$ & 0.0902 &   0.0852 &   0.1369\\
		$\beta$ &0.8702   & 0.8944   & 0.8059\\
			\hline
		\end{tabular}
	\end{center}
\end{table}	



\subsection{Evaluation of VaR and ES forecasts} \label{univariate_risk_empirical}

Assuming equally weighted portfolio returns, one-step-ahead daily VaR and ES forecasts from the proposed Semi-DCC-AL and Semi-DCC-FZ0 models are produced by using Equation (\ref{e:modvar_es_dcc}). In this section, these forecasts are compared with the ones from the competing models presented in Section \ref{data_empirical}.

The standard quantile loss function is employed to compare the models for VaR forecast accuracy: the most accurate VaR forecasts should minimize the quantile loss function, given as:
\begin{equation}\label{q_loss}
L_Q=\sum_{t=T_{\text{in}} +1}^{T_{\text{in}}+T_{\text{out}}} (r_t- \widehat{Q}_t) (\alpha-I(r_t \le \widehat{Q}_t)) \,\, ,
\end{equation}
where $T_{\text{in}}$ is the in-sample size, $T_{\text{out}}$ is the out-of-sample size and $\widehat{Q}_{T_{\text{in}}+1},\ldots,\widehat{Q}_{T_{\text{in}}+T_{\text{out}}}$ is a series of VaR forecasts at level $\alpha$ for observations $r_{T_{\text{in}}+1},$ $ \ldots,$ $r_{T_{\text{in}}+T_{\text{out}}}$.

Moving to the assessment of joint (VaR, ES) forecasts, as discussed in Section \ref{background_section}, \cite{tayl2019} shows that the negative logarithm of the likelihood function built from Equation (\ref{e:es_var_likelihood}) is strictly consistent for $Q_t$ and $\text{ES}_t$ considered jointly, and fits into the class of strictly consistent joint loss functions for VaR and ES developed by \cite{Fissler2016}. This loss function is also called the AL log-score in \cite{tayl2019} and is defined as:
\begin{equation}\label{es_caviar_log_score}
L_{\text{AL},t}(r_t, \widehat{Q}_t, \widehat{\text{ES}}_t) = -\text{log} \left( \frac{\alpha-1}{\widehat{\text{ES}}_t} \right) - {\frac{(r_t- \widehat{Q}_t)(\alpha-I(r_t\leq \widehat{Q}_t))}{\alpha \widehat{\text{ES}}_t}}.
\end{equation}
In our analysis, we first use the joint loss $L_{\text{AL}} = \sum_{t=T_{\text{in}}+1}^{T_{\text{in}}+T_{\text{out}}} L_{\text{AL},t}$ to formally and jointly assess and compare the VaR and ES forecasts from all models.

Meanwhile, the FZ0 loss used in \cite{pattonetal2019} can be calculated as:
\begin{equation}
L_{\text{FZ0},t}(r_t, \widehat{Q}_t, \widehat{\text{ES}}_t)=-\frac{1}{\alpha \text{ES}_t}I(r_t \le  Q_t) \,(Q_t-r_t)+\frac{Q_t}{\text{ES}_t}+\log(-\text{ES}_t)-1.
\label{e:fz0_loss}
\end{equation}

The FZ0 joint loss $L_{\text{FZ0}} = \sum_{t=T_{\text{in}}+1}^{T_{\text{in}}+T_{\text{out}}} L_{\text{FZ0},t}$ is also used to test the performance of the models in comparison.

\begin{figure}[!ht]
     \centering
\includegraphics[width=0.8\textwidth]{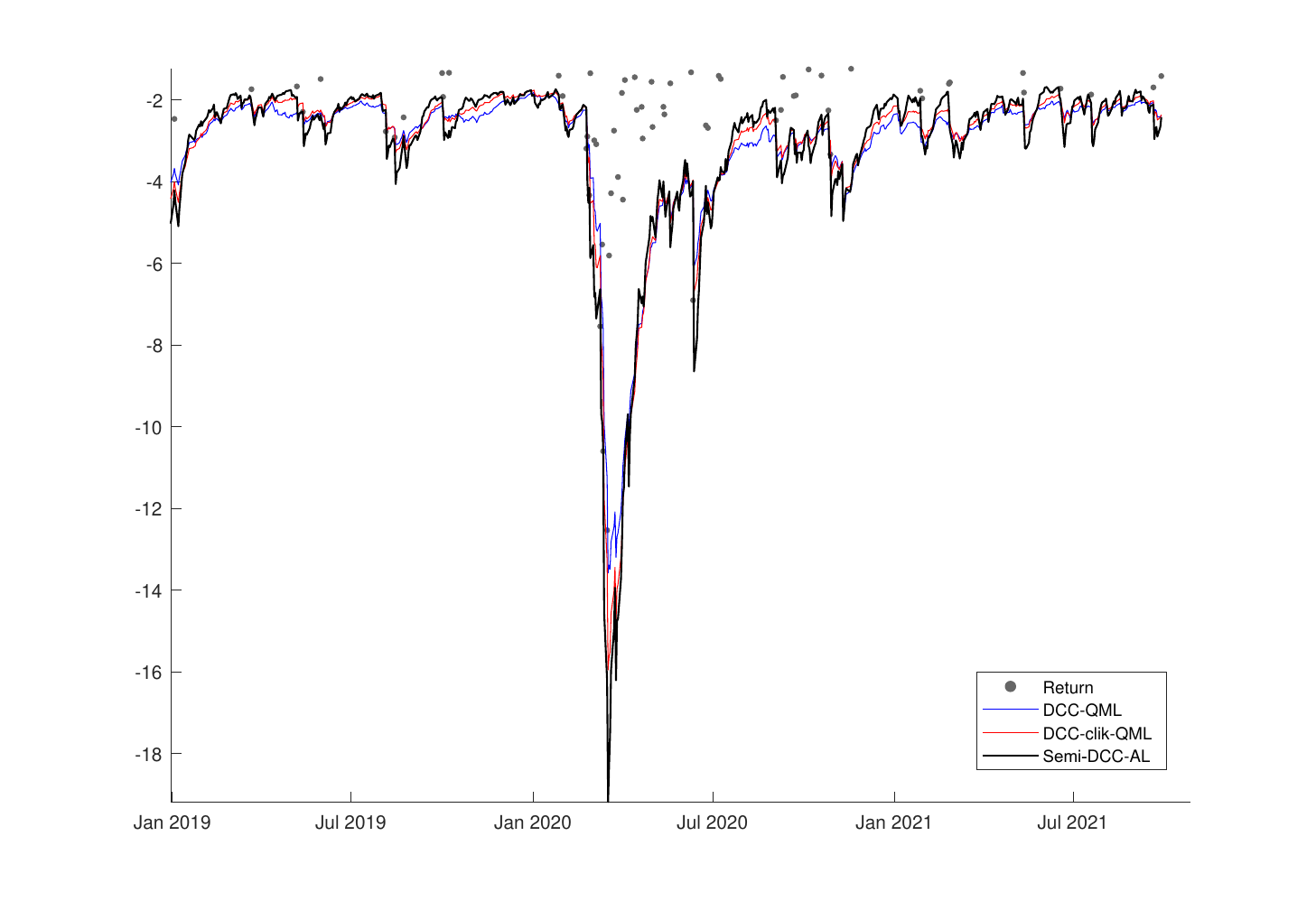}
\caption{\label{f:ES_forecasts_zoom_in} 2.5\% ES forecasts from the DCC-QML, DCC-clik-QML and Semi-DCC-AL models, using equally weighted Dow Jones returns.}
\end{figure}

Figure \ref{f:ES_forecasts_zoom_in} visualizes the 2.5\% portfolio ES forecasts from the DCC-QML, DCC-clik-QML and Semi-DCC-AL models for the Jan 2019--Sep 2021 period, using equally weighted Dow Jones returns. In general, we can see that the ES forecasts produced from the Semi-DCC-AL model have a long-run behaviour comparable to the ones from the DCC-QML and DCC-clik-QML. However, in the short term, forecasts from these models can be characterized by substantially different dynamic patterns. This is particularly evident in the period immediately following the outbreak of the COVID-19.

Inspecting the ES forecasts at the beginning of 2020, we can see that the three models have distinctive behaviours. Comparing to the DCC-QML model, the Semi-DCC-AL model is more reactive to the return shocks. This is because of the larger $a$ and $\alpha$ estimates as discussed in Tables \ref{t:corrcoeffs} and \ref{t:volcoeffs}. The ES forecasts from the DCC-clik-QML stay in between the ones from DCC-QML and Semi-DCC-AL, which is also consistent with the findings from Table \ref{t:corrcoeffs}.

The quantile and joint loss results from the proposed Semi-DCC-AL and Semi-DC-FZ0 models and other competing models are shown in Table \ref{t:risk_fore_results}. Overall, the results show that the proposed Semi-DCC-AL and Semi-DCC-FZ0 models, in comparison to the other models under analysis, generate competitive loss results, which lends evidence on the validity of the proposed semi-parametric framework. Furthermore, the Semi-DCC models generate smaller loss values than all other parametric or semi-parametric DCC models that have been taken as benchmarks. Meanwhile, the performance of Semi-DCC-AL and Semi-DCC-FZ0 is quite close to each other, showing that the choice of the joint loss has a minimal impact on the Semi-DCC estimation and risk forecasting performance.

The Model Confidence Set (MCS) of \cite{hansen2011_MCS} is employed to statistically compare the quantile loss (Equation (\ref{q_loss})), AL joint loss (Equation (\ref{es_caviar_log_score})) and FZ0 joint loss (Equation (\ref{e:fz0_loss})) values yielded by the different models. A MCS is a set of models constructed to contain the superior models with a given level of confidence, chosen in our paper as 75\%. The SQ method \citep[see][for details]{hansen2011_MCS} is employed to calculate the MCS test statistic. The MCS results, using both quantile and AL\&FZ0 joint loss functions, are also included in Table \ref{t:risk_fore_results}, with red highlighting indicates models that are excluded from the 75\% MCS. First, the AL and FZ0 loss function produce consistent MCS results as expected. Further, as can be seen the Semi-DCC-AL and Semi-DCC-FZ0 are always included in the MCS for the quantile test and both joint loss tests, across all probability levels. On the more extreme 1\% level, for the quantile loss test only CARE-IG is in the MCS in addition to the Semi-DCC. For the joint loss tests, most of the existing DCC models are not in the MCS, except DCC-clik-QML. On the 2.5\% level, all the DCC models, except the proposed Semi-DCC, are excluded from the MCS based on the joint loss tests. On the 5\% level, only the DCC-clik-N and Semi-DCC models are in the MCS based on joint loss.
\begin{table}[!ht]
\begin{center}
\scriptsize
\caption{\label{t:risk_fore_results} \small  The quantile loss (Equation (\ref{q_loss})), AL joint loss (Equation (\ref{es_caviar_log_score})) and FZ0 joint loss values (Equation (\ref{e:fz0_loss})) for all competing models across the full forecasting sample. Models in \cred{red} are excluded from the 75\% MCS according to the SQ statistic.\\}\tabcolsep=10pt
\begin{tabular}{lcccccc} \hline
Model& Quantile loss & AL Joint loss & FZ0 joint loss \\\hline
$\boldsymbol{\alpha=1\%}$ \\
CARE-SAV&	 \cred{43.8} 	& \cred{2,741.2} &	 \cred{1,491.4}   \\
CARE-IG	 &  39.3 	 &2,540.4 	& 1,294.9   \\
				
ES-CAViaR-SAV&	\cred{ 40.5} 	& \cred{2,576.7} 	 &\cred{1,331.4}   \\
ES-CAViaR-IG&  \cred{40.0} &	 2,547.8 	& 1,301.7 	 \\
				
DCC-QML&	 \cred{41.8} 	& \cred{2,628.8} &	 \cred{1,385.6}   \\
DCC-N	 &   \cred{43.8}& 	 \cred{2,720.3} 	 & \cred{1,472.1}  \\
DCC-clik-QML&	 \cred{40.6} 	& 2,585.5 	& 1,340.7    \\
DCC-clik-N	&  \cred{42.1} &	 \cred{2,679.5} &	 \cred{1,429.6}  \\
DCC-t	&   \cred{43.6} 	& \cred{2,695.1} 	& \cred{1,447.4}  \\
	
Semi-DCC-AL& 38.9 	& 2,528.5 	& 1,282.6 	 \\
Semi-DCC-FZ0 & 38.9 & 	 2,528.3 	& 1,282.5 	\\
\hline

$\boldsymbol{\alpha=2.5\%}$ \\

CARE-SAV&	 90.4 &	 2,405.4 & 1,133.8 \\
CARE-IG	 & 89.1 	& 2,392.1 & 1,120.4  \\
				
ES-CAViaR-SAV&	 89.4 &	 2,398.1 &  1,126.5 \\
ES-CAViaR-IG&	 87.3 	 &2,369.5 & 1,097.4 \\
				
DCC-QML&	 98.7 	& \cred{2,492.6}&  \cred{1,221.8} \\
DCC-N	 &\cred{101.2} &	 \cred{2,566.8} & \cred{ 1,290.8} \\
DCC-clik-QML&	 95.2 &	\cred{ 2,453.9}& \cred{1,181.4}  \\
DCC-clik-N	& 97.7 	& \cred{2,528.9} & \cred{1,251.3} \\
DCC-t	& \cred{101.5} &	 \cred{2,555.3}& \cred{1,279.8}  \\
	
Semi-DCC-AL&	 89.1 &	 2,386.8 &  1,115.0\\
Semi-DCC-FZ0 &	 89.0 &	 2,383.8 & 1112.0\\
\hline

$\boldsymbol{\alpha=5\%}$ \\	
CARE-SAV& 130.0 &	 2,070.2&	 764.9 	  \\
CARE-IG	 &  130.0 	 &2,070.7 	& 765.9   \\
				
ES-CAViaR-SAV&  131.3 	& 2,077.4 	 &770.1 \\
ES-CAViaR-IG&	 130.6 &	 2,065.1 	& 757.9  \\
				
DCC-QML& \cred{137.2}& 	\cred{ 2,157.3} &	 \cred{856.2} 	 \\
DCC-N	 &  136.2 	& \cred{2,143.9} &	 \cred{839.0}   \\
DCC-clik-QML&	 \cred{134.1} 	& \cred{2,123.3} 	& \cred{820.1}    \\
DCC-clik-N	&  132.7 	& 2,108.4 	 &801.2  \\
DCC-t	&  \cred{136.5} 	& \cred{2,138.5} 	& \cred{833.4}  \\
	
Semi-DCC-AL& 131.1 &	 2,079.9 &	 772.9 	 \\
Semi-DCC-FZ0 & 131.1 	& 2,083.9 	 &776.7 	 \\
\hline
\end{tabular}
\end{center}

\end{table}

To further backtest the VaR forecasts, we have employed several quantile accuracy and independence tests, including: the unconditional coverage (UC) test \citep{kupiec1995techniques}; the conditional coverage (CC) test \citep{christoffersen1998evaluating}; the dynamic quantile (DQ) test \citep{engle2004caviar}; and the quantile regression based VaR (VQR) test \citep{gaglianone2011evaluating}. Table \ref{var_backtest} presents the UC, CC, DQ (lag 1 and lag 4) and VQR backtests' p-values for the 11 competing models. The ``Total'' columns show the total number of rejections at the 5\% significance level. As can be seen, across the 1\% and 2.5\% probability levels the Semi-DCC is the only framework which receives 0 rejections. On the 5\% level, the Semi-DCC is rejected 3 times, while all the other models are rejected by all the 5 tests. The DCC-N and DCC-t models are more likely to be rejected by various tests. This demonstrates the importance of the return distribution selection in parametric DCC models. The Semi-DCC model, which is semi-parametric, shows advantage from this perspective. Comparing the Semi-DCC model to the other semi-parametric DCC models with QML, the Semi-DCC still has better performance considering these backtests. These results again lend evidence on the validity and effectiveness of the Semi-DCC in VaR forecasting.

\begin{table}[!ht]
\begin{center}
\scriptsize
\caption{\label{var_backtest}Summary of VaR forecasts UC, CC, DQ1, DQ4, and VQR backtests' p-values and the total number of rejections at the 5\% significance level.}\tabcolsep=10pt
\begin{tabular}{lccccc|c} \hline

Model&	UC	&CC&	DQ1&	DQ4&	VQR & Total\\ \hline

$\boldsymbol{\alpha=1 \%}$ \\
CARE-SAV&2\%&6\%&23\%&63\%&1\%&2\\
CARE-IG&34\%&60\%&78\%&98\%&64\%&0\\
ES-CAViaR-SAV&20\%&42\%&69\%&4\%&76\%&1\\
ES-CAViaR-IG&34\%&60\%&78\%&98\%&60\%&0\\
DCC-QML&53\%&15\%&3\%&0\%&82\%&2\\
DCC-N&19\%&3\%&0\%&0\%&6\%&3\\
DCC-clik-QML&97\%&2\%&0\%&0\%&89\%&3\\
DCC-clik-N&7\%&2\%&0\%&0\%&0\%&4\\
DCC-t&19\%&3\%&0\%&0\%&5\%&3\\
Semi-DCC-AL&20\%&42\%&67\%&95\%&78\%&0\\
Semi-DCC-FZ0&20\%&42\%&67\%&95\%&65\%&0\\
\hline
$\boldsymbol{\alpha=2.5\%}$ \\

CARE-SAV&	76\%&	16\%&	9\%&	3\%&	46\%&	1\\
CARE-IG	&95\%&	46\%&	41\%&	6\%	&0\%&	1\\
ES-CAViaR-SAV&	90\%&	14\%&	8\%&	2\%&	68\%&	1\\
ES-CAViaR-IG&	67\%	&83\%&	82\%&	1\%&	84\%&	1\\

DCC-QML	&76\%	&3\%&	0\%&	0\%&	0\%&	4\\
DCC-N&	4\%	&2\%&	0\%&	0\%&	0\%&	5\\
DCC-clik-QML&	51\%&	18\%&	6\%&	0\%&	20\%&	1\\
DCC-clik-N&	9\%&	10\%&	4\%&	0\%&	5\%&	2\\
DCC-t&	13\%&	4\%&	0\%&	0\%&	11\%&	3\\
Semi-DCC-AL&	67\%&	83\%&	81\%&	8\%&	42\%&	0\\
Semi-DCC-FZ0 & 81\%	&91\%	&80\%&	11\%	&40\%& 0\\
\hline
$\boldsymbol{\alpha= 5 \%}$ \\

CARE-SAV&0\%&1\%&3\%&1\%&2\%&5\\
CARE-IG&0\%&1\%&3\%&1\%&2\%&5\\
ES-CAViaR-SAV&0\%&0\%&1\%&0\%&1\%&5\\
ES-CAViaR-IG&0\%&0\%&2\%&1\%&3\%&5\\
DCC-QML&0\%&0\%&0\%&0\%&0\%&5\\
DCC-N&1\%&1\%&1\%&0\%&0\%&5\\
DCC-clik-QML&0\%&0\%&1\%&0\%&0\%&5\\
DCC-clik-N&1\%&2\%&4\%&0\%&0\%&5\\
DCC-t&1\%&1\%&1\%&0\%&0\%&5\\
Semi-DCC-AL&1\%&2\%&7\%&0\%&12\%&3\\
Semi-DCC-FZ0&1\%&2\%&7\%&0\%&5\%&3\\

\hline
\end{tabular}
\end{center}
\end{table}

Lastly, \cite{bayer2022regression} propose three versions of ES backtests named as Auxiliary, Strict and Intercept  ES regression (ESR) backtests\footnote{For the implementation, we use the R package developed by the authors, which can be found at:  \url{https://search.r-project.org/CRAN/refmans/esback/html/esr_backtest.html}. The link also includes the details of the three versions of the backtests.}. These ESR backtests (two-sided), labeled as A, S, and I respectively, are also employed to backtest the ES forecasts from the 11 competing models.


As in Table \ref{es_regression_backtest}, the three backtests return quite consistent results, with the Semi-DCC models being not rejected. On the 2.5\% probability level the models that get rejected by all three backtests are DCC-N, DCC-clik-N and DCC-t. On the 1\% and 5\% levels, the DCC-N and DCC-QML are rejected twice, respectively.

\begin{table}[!ht]
\begin{center}
\scriptsize
\caption{\label{es_regression_backtest}Summary of three versions (A, S, I) of ES backtests' p-values and the total number of rejections at the 5\% significance level.}\tabcolsep=10pt
\begin{tabular}{lccc|c} \hline

Model &	A &	S &	I &	Total \\ \hline
$\boldsymbol{\alpha=1 \%}$ \\
CARE-SAV&22\%&22\%&21\%&0\\
CARE-IG&100\%&100\%&87\%&0\\
ES-CAViaR-SAV&89\%&89\%&86\%&0\\
ES-CAViaR-IG&96\%&96\%&75\%&0\\
DCC-QML&65\%&64\%&94\%&0\\
DCC-N&2\%&2\%&7\%&2\\
DCC-clik-QML&94\%&94\%&97\%&0\\
DCC-clik-N&9\%&9\%&21\%&0\\
DCC-t&10\%&10\%&7\%&0\\
Semi-DCC-AL&99\%&99\%&87\%&0\\
Semi-DCC-FZ0&99\%&99\%&86\%&0\\
\hline
$\boldsymbol{\alpha=2.5 \%}$ \\

CARE-SAV&	83\%&	83\%&	70\%&	0\\
CARE-IG	&95\%&	95\%&	76\%&	0\\

ES-CAViaR-Mult-SAV&	92\%&	92\%&	78\%&	0\\
ES-CAViaR-IG&	96\%&	96\%&	96\%&	0\\

DCC-QML	&20\%&	20\%&	24\%&	0\\
DCC-N&	0\%	&0\%&	1\%	&3\\

DCC-clik-QML&	35\%&35\%&	21\%&	0\\
DCC-clik-N&	0\%	&0\%&	1\%	&3\\
DCC-t&	2\%	&2\%&	1\%	&3\\
Semi-DCC-AL	&87\%&	87\%&	71\%&	0\\
Semi-DCC-FZ0	&85\%&	85\%&	71\%&	0\\
\hline
$\boldsymbol{\alpha=5 \%}$ \\
CARE-SAV&18\%&18\%&12\%&0\\
CARE-IG&16\%&16\%&10\%&0\\
ES-CAViaR-SAV&17\%&17\%&12\%&0\\
ES-CAViaR-IG&14\%&14\%&10\%&0\\
DCC-QML&2\%&2\%&17\%&2\\
DCC-N&27\%&27\%&41\%&0\\
DCC-clik-QML&11\%&12\%&19\%&0\\
DCC-clik-N&66\%&65\%&46\%&0\\
DCC-t&38\%&38\%&52\%&0\\
Semi-DCC-AL&43\%&44\%&21\%&0\\
Semi-DCC-FZ0&47\%&47\%&25\%&0\\

\hline
\end{tabular}
\end{center}
\end{table}

Lastly, comparing to the univariate semi-parametric risk forecasting models, such as ES-CAViaR or CARE, Semi-DCC models can be used for asset allocation. Therefore, in Appendix \ref{po_empirical} we evaluate the out-of-sample performance of the Semi-DCC model in computing global minimum ES portfolios at the 2.5\% level. Further, we compare the properties of the minimum ES portfolios obtained by Semi-DCC with those obtained by standard DCC models which are used as benchmarks. The effectiveness of the proposed Semi-DCC framework in portfolio optimization is demonstrated.

\section{Conclusions}\label{conclusion_section}

In this paper, we propose an innovative framework for forecasting portfolio tail risk in a multivariate and semi-parametric setting. The proposed framework is capable of modeling high dimensional return series parsimoniously and efficiently. Compared to the state-of-the-art univariate semi-parametric models, our approach delivers competitive risk forecasting performances and, in addition, it can be used for a wider range of applications including asset allocation. Compared to existing DCC estimation approaches, our empirical application shows that the Semi-DCC is able to deliver more accurate risk forecasts. Shifting the focus to the computation of minimum ES portfolios, we find that the Semi-DCC produces competitive performances compared to other DCC models, especially for the highly volatile ``COVID 19'' period.

Potential projects for future research include a deeper investigation of the performance of the Semi-DCC model in asset allocation, as well as an extension of the proposed framework to other settings such as the prediction of CoVaR \citep{AdrBru2016}, building on recent results in semi-parametric estimation by \cite{FisHog2024} and \cite{DimHog2024}.

\cite{gerlach2020semi} extend the framework in \cite{tayl2019} by incorporating realized measures as exogenous variables, showing improved VaR and ES forecast accuracy. As our analysis in this paper is based on low-frequency daily data, our research plans also include extending the Semi-DCC model by incorporating high-frequency realized measures as exogenous variables to further improve its performance.

Furthermore, extending the Semi-DCC framework to more complex parameterizations, such as the AG-DCC model, is an interesting direction for future research. In this regard, it is worth noting that shifting the focus to alternative parameterizations could also pave the way for the investigation of alternative empirical frameworks. For example, a variant of the AG-DCC has recently been used by \cite{pineda2022} to analyze financial contagion.

Finally, inspired by the results of our asset allocation exercise, we expect that Semi-DCC models could be well suited for tail risk hedging applications, especially when the hedged assets exhibit extreme volatility dynamics. However, despite the empirical relevance of this topic, its investigation goes beyond the scope of this paper and is currently left for future research.

\section*{Disclosure of Interest}
No conflict of interests to be declared.

\section*{Disclosure of Funding}
Giuseppe Storti acknowledges financial support under the National Recovery and Resilience Plan (NRRP), Mission 4, Component 2, Investment 1.1, Call for tender No. 104 published on 2.2.2022 by the Italian Ministry of University and Research (MUR), funded by the European Union – NextGenerationEU– Project Title: Methodological and computational issues in large-scale time series models for economics and finance (ID: 20223725WE) – CUP  D53D 2300610 0006 - Grant Assignment Decree No. n. 967 of 30/06/2023.

\section*{Data Availability Statement}
Data were downloaded from https://finance.yahoo.com/. The authors confirm that the data supporting the findings of this study are available within the supplementary materials of the paper.

\clearpage

\bibliographystyle{chicago}
\bibliography{bibliography}

\clearpage
\appendixtitleon
\appendixtitletocon
\begin{appendices}

{\centering
\section{Minimum ES portfolios}\label{po_empirical}
\par
}

We compare the properties of the 2.5\% minimum ES portfolios obtained by Semi-DCC-AL with those obtained by standard DCC models, which are used here as benchmarks. Our interest for the computation of minimum ES portfolios is motivated by the prominent role that this risk measure is assuming in the international regulatory framework.

In order to compute minimum ES portfolios, we develop a two-stage procedure. At the initial stage, an estimate of the conditional variance and covariance matrix $H_t$ is obtained by fitting a Semi-DCC-AL model to an equally weighted portfolios. The optimal portfolio weights are then identified by minimizing the portfolio ES conditional on the first stage volatility estimates. There are two main reasons for our preference for this type of approach: first, it is computationally convenient as it avoids re-estimating the model at each iteration, and second, it can be easily implemented for semi-parametric models other than the Semi-DCC-AL model, thus facilitating comparison with benchmark DCC models.

The main steps of the procedure are summarised below:

\begin{enumerate}
	\item Given an initial set of portfolio weights $\mathbf{w}$, fit a Semi-DCC-AL (or other standard DCC models) and save $\hat{H}_t$ for each in-sample time stamp $t$
        \[
	\hat{H}_t= \hat{D}_t \hat{P}_t \hat{D}_t, \qquad t=1,\ldots,T.
	\]
	\item Compute standardized portfolio returns as below
	\[
	\mathbf{z}_t=\frac{\mathbf{w}'(\mathbf{r}_t-\hat{\boldsymbol{\mu}})}{(\mathbf{w}'H_t\mathbf{w})^{1/2}}\qquad t=1,\ldots,T,
	\]
 where $\hat{\boldsymbol{\mu}}$ is the mean vector of the portfolio returns.
	\item Estimate the sample quantiles $\hat{q}_{\alpha}(\mathbf{w})$ and sample tail expectation $\hat{c}_{\alpha}(\mathbf{w})$ of $\mathbf{z}_t$.
	\item Produce the portfolio ES forecasts as functions of $\mathbf{w}$ 
	\begin{align*}
		\widehat{\text{ES}}_{T+1}(\mathbf{w}) &=\mathbf{w}'\hat{\boldsymbol{\mu}}+\hat{c}_{\alpha}(\mathbf{w}) (\mathbf{w}'\hat{H}_{T+1}\mathbf{w})^{1/2}
	\end{align*}
\item Minimum ES portfolios are then calculated by optimizing $\widehat{\text{ES}}_{T+1}(\mathbf{w})$ with respect to $\mathbf{w}$ under the set of convex constraints:
 \[  0 \le w_i \le 1; \qquad
 \sum_{i=1}^{n} w_i= 1,
 \]



\end{enumerate}

It is worth noting that although our settings exclude short selling by restricting portfolio weights to values in [0,1], removing this restriction is a trivial and immediate extension of the procedure. Minimization has been performed using the Matlab ``fmincon'' optimization routine in Matlab 2023a.

As benchmarks for comparison we consider both semi-parametric (DCC-QML and DCC-clik-QML) and parametric (DCC-t) models. An equally weighted (EW) portfolio is also included as a further benchmark. The portfolio allocation exercise is based on the same rolling window forecasting scheme described in Section \ref{univariate_risk_empirical}, with an in-sample size of $T_{\text{in}}=3000$  and out-of-sample size $T_{\text{out}}=1213$ for the same 28 assets from the Dow Jones index.

\begin{figure}[htb]
	\includegraphics[width=\textwidth]{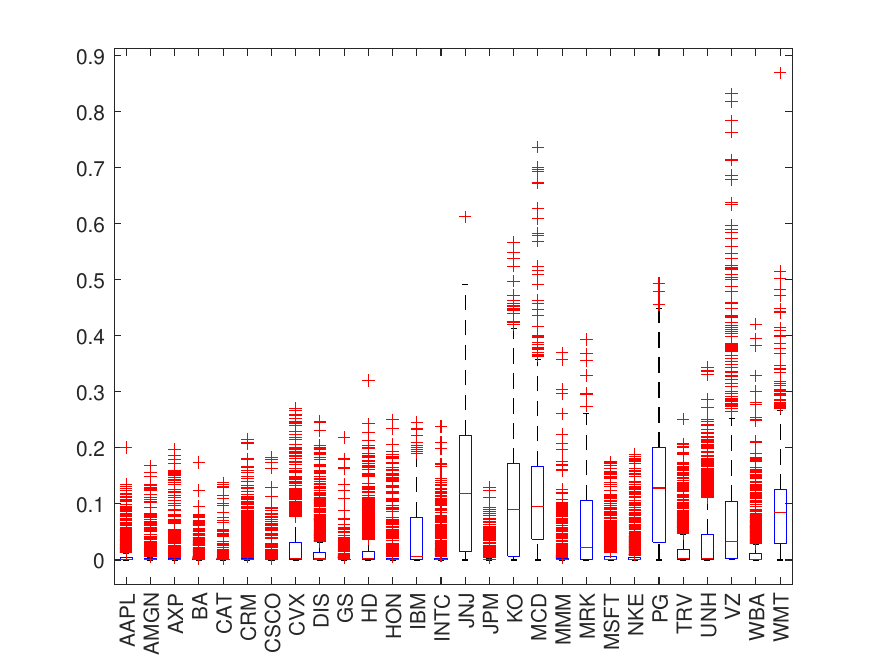}
	\caption{\label{f:es_po_weight} Boxplot of the Semi-DCC-AL 2.5\% minimum ES portfolio weights for all forecasting steps.}
\end{figure}

For all models out-of-sample 2.5\% minimum ES portfolios are computed. Figure \ref{f:es_po_weight} illustrates the Semi-DCC-AL optimized minimum ES portfolio weights for all the forecasting steps.  Table \ref{t:po_forecast} compares the statistical properties of the minimum ES portfolio obtained by the Semi-DCC-AL model with the benchmarks. Firstly, we focus on the value of the empirical ES of optimized portfolios which is expected to be higher (closer to 0) for more effective risk minimization strategies. Compared to the EW portfolio, all estimated DCC models yield substantial reductions in terms of out-of-sample empirical ES. Meanwhile, Semi-DCC-AL and the other DCC models under analysis produce quite similar  performances for the whole-out-of-sample period, with Semi-DCC-AL and DCC-clik-QML slightly outperforming the other DCC models. Such results lend evidence on the general validity of the proposed Semi-DCC-AL for asset allocation purposes. Regarding other descriptive statistics of portfolio returns, the Semi-DCC-AL produces smallest kurtosis and is least skewed. It also has the most conservative minimum value. These results provide evidence that the returns of the minimum ES portfolio obtained by the Semi-DCC-AL model are characterized by lighter tails and a less elongated left tail compared to the benchmarks, leading to a lower probability of occurrence of negative extreme events.

\begin{table}[ht!]
\caption{\label{t:po_forecast} 2.5\% minimum ES portfolio returns under different portfolio configurations produced by various models. Portfolio descriptive statistics include: $\widehat{ES}_\alpha$: sample $\alpha$-level ES ; $\kappa$: sample kurtosis; $skew$: sample skewness; $min$: sample minimum.}	
\begin{center}
\begin{tabular}{lccccc}
& DCC-QML & DCC-clik-QML & DCC-t & EW & Semi-DCC-AL \\
\hline
$\widehat{ES}_{\alpha}$ & -3.555 & -3.496 & -3.513 & -4.195 & -3.495 \\
$\kappa$ & 23.29 & 18.618 & 25.935 & 26.668 & 16.308 \\
$skew$ & -1.404 & -0.926 & -1.572 & -0.987 & -0.587 \\
$min$ & -10.43 & -8.119 & -10.934 & -12.533 & -7.729 \\
\hline
\end{tabular}

\end{center}

\end{table}

\begin{figure}[htb]
	\includegraphics[width=\textwidth]{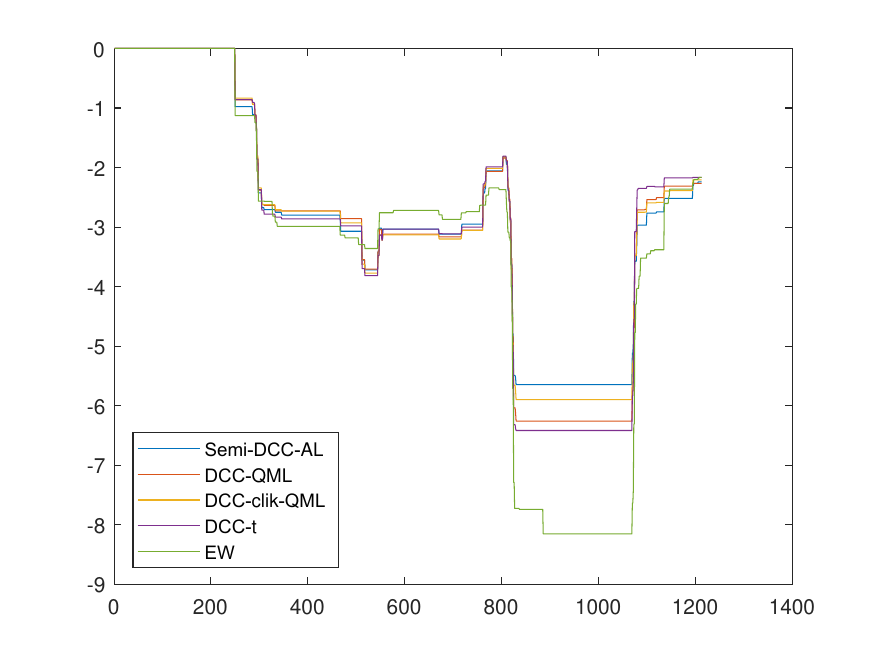}
	\caption{\label{f:ma_es_forecast} Moving 2.5\% ES (sample tail mean) of optimized portfolio returns.}
\end{figure}

In addition, to examine the differences in the performance of the different models over a shorter time horizon, Figure \ref{f:ma_es_forecast} shows the moving empirical tail-mean (2.5\% ES) of the optimised portfolio returns computed over a 250-day annual window. It is of particular interest to focus on the ``COVID-19'' period, which coincides with the negative peak in the plot, approximately centred on day 1000. First, we see that, in this highly risky period, the EW allocation is not effective in risk diversification since it yields remarkably lower ES values. At the same time, the DCC-t model, which generates competitive portfolio ES for the entire forecast period, is outperformed by all other semi-parametric DCC-type models for the high volatility ``COVID-19'' period, demonstrating the effectiveness of the semi-parametric approach, which is free of return distribution selection. Finally, focusing on the semi-parametric methods, it is evident that Semi-DCC-AL is able to produce preferred portfolio ES forecasts compared to DCC-QML and DCC-clik-QML for the ``COVID-19'' period.

\end{appendices}
\end{document}